\documentclass[journal=jctcce,manuscript=article,layout=twocolumn]{achemso}

\usepackage{graphicx}
\usepackage{dcolumn}
\usepackage{bm}

\usepackage[utf8]{inputenc}
\usepackage[T1]{fontenc}
\usepackage{mathptmx}
\usepackage{comment}
\usepackage{etoolbox}
\usepackage{amsmath}
\usepackage{bm}
\usepackage{hyperref}
\usepackage{cleveref}
\usepackage{rotating}
\usepackage{dsfont}
\usepackage{booktabs}
\usepackage{notes2bib}
\usepackage{braket}
\usepackage{xcolor}
\usepackage{siunitx}
\usepackage{array}
\usepackage{threeparttable}

\usepackage[section]{placeins}

\def\br{\mathbf{r}}

\usepackage{cuted}

\def\h2o{\mathrm{H}_2\mathrm{O}}

\newcommand{\nn}{\nonumber}

\DeclareUnicodeCharacter{2212}{\ensuremath{-}} \author{Paul van Hoegaerden}
\affiliation{Department of Chemistry and Pharmaceutical Sciences, Vrije Universiteit, De Boelelaan 1105, 1081 HV Amsterdam, The Netherlands}
\author{Arno F\"orster}
\email{a.t.l.foerster@vu.nl}
\affiliation{Department of Chemistry and Pharmaceutical Sciences, Vrije Universiteit, De Boelelaan 1105, 1081 HV Amsterdam, The Netherlands}

\title{Second-order Bethe--Salpeter kernels: the interplay of vertex-corrected quasiparticle energies and the screened interaction}

\begin{document}

\begin{abstract}
We study second-order Bethe–Salpeter kernels generated by truncating an iterative formulation of Hedin's equations after the second cycle. Starting from the Hartree--Fock or the $GW$ self-energy, whose first cycles correspond to TDHF and $GW$+BSE, we obtain the second-cycle kernel by differentiating the first-cycle correlation self-energy with the interacting correlation function replaced by the non-interacting one, and evaluate it in the static limit of the effective kernel. Using a fixed restricted Hartree--Fock orbital basis, we benchmark 30 singlet and 28 triplet excitation energies for seven small molecules (six for the triplets) against the theoretical best estimates from QUEST v1.2.
The second cycle lowers the mean absolute error (MAE) in both sequences. For the $GW$ sequence, the MAE decreases from 0.63 to 0.29~eV (singlets), on par with ADC(2) (0.30~eV), and from 0.43 to 0.38~eV (triplets). BSE$^{(2)}$@HF removes the triplet instabilities of TDHF and yields MAEs of 0.43 and 0.38~eV, or 0.31 and 0.30~eV when the Tamm--Dancoff approximation is applied in all cycles. Assigning the statically screened interaction $W_0$ to the exchange part of the $GW$ self-energy counts second-order screening twice in the second-cycle kernel and strongly blueshifts valence excitations. Without this double counting, dressing the exchange-type terms with $W_0$ lowers the singlet MAE from 0.37 to 0.29 eV but leaves the triplets nearly unchanged. Replacing second-order Green's function (GF2) quasiparticle energies by vertex-corrected ones in the same kernel raises excitation energies by about 0.7~eV, improving Rydberg but degrading valence excitations. A perturbative dynamical correction of the kernel, nearly independent of the quasiparticle energies, blueshifts valence excitations by 0.3--0.4~eV and increases the errors of all full-matrix BSE$^{(2)}$ variants. While ADC(2) remains cheaper and more accurate overall, the consistent improvement over the first cycle suggests that higher cycles are worth pursuing.
\end{abstract}

\section{Introduction}
\label{sec:intro}

Accurate first-principles predictions of neutral excitation energies remain a central challenge in computational spectroscopy.~\cite{Loos2020f, Loos2025QUEST} Green's-function many-body perturbation theory provides a systematic, diagrammatic framework for describing both charged and neutral excitations at a more favorable computational cost than high-level coupled-cluster methods.~\cite{Blase2020,Reining2018,martin2016} Whether it can also reach the same accuracy for molecules is a matter of active research. Its most widely used approximation, $GW$, was introduced by Hedin as an approximation to the electronic self-energy $\Sigma$.~\cite{Hedin1965} Here, $G$ denotes the one-particle Green's function and $W$ the dynamically screened Coulomb interaction, usually evaluated in the random-phase approximation (RPA).~\cite{Macke1950,Bohm1953,Golze2019} $GW$ provides quasiparticle (QP) energies, the highest occupied and lowest unoccupied of which are the negatives of the ionization potential (IP) and the electron affinity (EA), respectively. Combining $GW$ with the Bethe--Salpeter equation (BSE) yields neutral excitation energies that incorporate the direct attraction between an excited electron and the hole it leaves behind.~\cite{Strinati1988,Rohlfing2000,Blase2018,Blase2020} At its best, when combined with partial self-consistency, $GW$+BSE has been reported to yield typical errors of 0.1--0.3~eV for molecular singlet excitations and a mean error of approximately 0.2~eV on standard sets.~\cite{Jacquemin2015, Bruneval2015, Gui2018a, Blase2018, Blase2020, Loos2021}

The success of $GW$ is, in a sense, surprising. Hedin's exact equations couple the self-energy to a vertex function $\Gamma$, which $GW$ replaces by unity, discarding all vertex corrections.\cite{Hedin1965} That such a drastic simplification yields quantitatively useful QP energies has long been attributed to a cancellation between vertex corrections in the self-energy and in the screened interaction,\cite{Mahan1989, Bobbert1994, delSole1994, Bruneval2005} an interpretation that has recently received quantitative support.~\cite{Forster2024} The vertex is not the only aspect in which practical $GW$ departs from the exact theory. In practice, the Dyson equation is rarely solved self-consistently, leaving the resulting QP energies dependent on the Green's function used as the starting point. Imposing self-consistency removes this dependence and improves upon $G_0W_0$ built on Hartree--Fock or (semi)local density-functional starting points.~\cite{Caruso2012} However, increasing the degree of self-consistency does not systematically improve the QP energies. A starting point containing a substantial fraction of exact exchange instead acts as an inexpensive surrogate for self-consistency, giving at least as accurate molecular IPs at a fraction of the cost.~\cite{Bruneval2013,Bruneval2021a, McKeon2022}

Explicit vertex corrections have instead given mixed results. Including the vertex in the self-energy alone, a construction usually denoted as $G3W2$ or approximations to it, systematically increases IPs compared to $GW$ and therefore only improves results for Kohn-Sham starting points that are known to give systematically underestimated IPs.~\cite{Ren2015, Bruneval2024} Correcting the polarizability alone fails in the same way, in the opposite direction.~\cite{Lewis2019, Bruneval2021a} The lesson is that including vertex corrections does not, by itself, guarantee an improvement over $GW$. Systematic improvement for molecules requires mutually consistent vertices in both the self-energy and screened interaction.~\cite{Forster2024}

The same issue arises for neutral excitations. The BSE kernel is formally the functional derivative of the self-energy with respect to the one-particle Green's function, $\Xi=i\delta\Sigma/\delta G$. For $\Sigma=iGW$, this derivative generates both a screened-exchange contribution and terms involving $\delta W/\delta G$. In standard BSE calculations, the latter are neglected and the screened interaction is evaluated in the static limit, $W(\omega)\rightarrow \mbox{$W(\omega=0)$}\equiv W_0$.~\cite{Hanke1979, Hanke1980, Strinati1988, Rohlfing2000,Onida2002,Blase2018} The static limit renders the kernel instantaneous and reduces the BSE to a linear eigenvalue problem, thereby restricting the number of solutions to the number of particle--hole pairs and excluding states with appreciable double-excitation character.~\cite{Zhang2013b,Romaniello2009b,Sangalli2011,Authier2021,Bintrim2022} Its validity rests on the excitonic binding energy being small compared with the plasmon frequencies that govern the dynamics of $W$, a condition met in many semiconductors but not generally in molecules.~\cite{Rohlfing2000} Restoring the frequency dependence through explicit upfolding~\cite{Bintrim2022} or a self-consistent solution of the frequency-dependent problem can recover such states, whereas a perturbative treatment~\cite{Ma2009,Ma2010,Loos2020b} only corrects the existing single-excitation roots. Neglecting $\delta W/\delta G$ has consequences of a different kind. Numerically, the discarded terms are at most $0.2$~eV for the valence states of Thiel's set~\cite{Schreiber2008} and partly cancel among themselves.~\cite{Yamada2022,Yamada2023} Monino and Loos found that including them does not improve on BSE@$G_0W_0$@HF.~\cite{Monino2023} However, they reach up to $0.8$ eV for intramolecular charge-transfer excitations, where they exceed the first-order $(ia|jb)$ term.~\cite{Yamada2023} More fundamentally, their omission breaks the formal connection between the kernel and the self-energy that generated the propagator.

This connection matters because it underlies the conserving character of a self-consistent approximation. Luttinger and Ward showed that the grand potential can be written as a stationary functional of $G$.~\cite{Luttinger1960a} The interaction contribution is contained in $\Phi[G]$, which is the sum of closed skeleton diagrams and whose functional derivative gives the self-energy, $\Sigma=\delta\Phi/\delta G$. Differentiating the Dyson equation with respect to an external perturbation $U$ yields the BSE for the two-particle correlation function $L\propto\delta G/\delta U$, with the kernel defined as $\Xi=i\delta\Sigma/\delta G=i\delta^2\Phi/\delta G\,\delta G$. If $\Sigma$ is $\Phi$-derivable, $G$ is self-consistent, and $\Xi$ is the derivative of this same $\Sigma$, the BSE response is conserving in the Baym--Kadanoff sense and satisfies the associated sum rules.~\cite{Baym1961,Baym1962} In practice, however, most $GW$+BSE calculations do not meet these conditions simultaneously. Practical $GW$ implementations are usually not fully self-consistent,~\cite{Stan2009} and the conventional BSE kernel, for the reasons given above, is not the derivative of the self-energy used to dress $G$. Standard $GW$+BSE is therefore not a conserving approximation. Hartree--Fock (HF) provides a useful contrast. Since $\Sigma_{\mathrm{HF}}$ is $\Phi$-derivable and does not depend on $W$, its kernel follows without approximation and is instantaneous. The corresponding first-cycle response, time-dependent HF (TDHF), is therefore conserving.

Conservation laws alone, however, do not ensure accurate individual excitation energies. The TDHF kernel is first order in the bare interaction $v$ and thus lacks screening of the electron--hole interaction, which enters only at higher order. Replacing $v$ by $W$ in the BSE kernel restores this missing physics and improves the excitation energies, but it also disconnects the kernel from the self-energy used to dress the propagator. Thus, internal consistency in the derivation of the kernel and self-energy is necessary for a conserving approximation, but is not by itself sufficient for quantitative accuracy.

A separate limitation concerns the crossing symmetry of the two-particle vertex. The Pauli principle requires the full vertex to be antisymmetric under the exchange of two fermion lines.~\cite{Bickers2004} The vertex generated by the $GW$ kernel violates this symmetry already at second order in $v$.~\cite{Krien2021} Parquet approaches enforce crossing symmetry, but, except for the exact solution, it cannot be satisfied simultaneously with the conserving property.\cite{Smith1992,Kugler2018} Since standard $GW$+BSE is neither conserving nor crossing symmetric, its accuracy appears to rely instead on a delicate balance between the screening in the kernel and the QP energies, as in $GW$ calculations.~\cite{Mahan1989,Bobbert1994,delSole1994,Forster2024}

These considerations motivate the central question of this work: how does the interplay between the QP energies and the interaction entering the second-order BSE kernel affect the accuracy of neutral excitation energies? The expectation is natural. More accurate QP energies should yield more accurate QP energy differences in the kernel, and thereby more accurate excitation energies. Improving both together, however, need not preserve the balance on which the conventional construction relies.

An alternative to Hedin's equations, that is often encountered in the recent literature~\cite{Romaniello2012, Starke2012, Forster2024, Forster2025, Vacondio2024, Patterson2024, Orlando2023} and goes back to the work of Baym and Kadanoff~\cite{Baym1961,Baym1962}, provides a convenient framework in which the self-energy and the neutral-excitation kernel can be treated on the same footing.~\cite{martin2016} In this formulation the screened interaction never appears as a separate quantity. The equations instead couple the bare Coulomb interaction $v$, the two-particle correlation function $L$, the Green's function $G$, the self-energy $\Sigma$, and the kernel $\Xi = i\,\delta\Sigma/\delta G$. Screening is generated through the repeated solution of the Dyson-like equation for $L$. Iteration consequently generates a sequence of approximations in which the self-energy and kernel are derived from one another at each step. The first cycle of this sequence has recently been investigated in detail.~\cite{Forster2024,Forster2025,Vacondio2024,Patterson2024} Here we extend the construction to the second cycle, taking as input the QP energies of the vertex-corrected self-energies of Ref.~\citenum{Forster2024} and starting from either an HF or a $GW$ self-energy, which yield BSE$^{(2)}$@HF and BSE$^{(2)}$@$GW$. To keep the cost at that of second-order methods, we replace the interacting $L$ of the first cycle by its non-interacting counterpart before differentiating. For the HF starting point, the kernel coincides, before the static limit is taken, with the dynamical second-order kernel of Ref.~\citenum{Zhang2013b}, and, in effective-kernel form, with that of Ref.~\citenum{Rebolini2016}. For the $GW$ starting point, $W_0$ additionally dresses one interaction line of the exchange-type terms.

The remainder of the paper is organized as follows. Section~\ref{sec:theory} introduces the iterative equations, starting points, second-cycle kernel, and its static and dynamical treatments. We then describe the computational details in Sec.~\ref{sec:CompDetails}. Section~\ref{sec:results} presents the benchmark results, separates the effects of the screened interaction and the QP energies, and examines the dynamical correction, the coupling block, and the particle--particle channel. We conclude in Sec.~\ref{sec:conclusions} by discussing the approximations of the present scheme and the extensions they suggest. %
\section{Theory} \label{sec:theory}

We use the formulation of Hedin's equations outlined in the Introduction. Our notation follows closely that of Ref.~\citenum{Schindlmayr1998}. The iterative cycle is seeded by an initial self-energy and Green's function, denoted by $\Sigma^{(0)}$ and $G^{(0)}$, respectively. In the first cycle, the quantities are obtained in the order
\[
    \Xi^{(1)} \rightarrow L^{(1)} \rightarrow \Sigma^{(1)}
    \rightarrow G^{(1)} ,
\]
after which the cycle is repeated with $G^{(1)}$ replacing $G^{(0)}$. With integration over repeated indices implied throughout, the equations at iteration $n$ are
\begin{align}
    &\Xi^{(n)}(1234) = i\, \frac{\delta \Sigma^{(n-1)}(13)}{\delta G^{(n-1)}(42)} \,, \label{eq:nthorderHedin1}\\
    &L^{(n)}(1234) = L^{(n)}_0(1234)
        \nn \\
        &\qquad \qquad \quad +  L^{(n)}_0(1536)\,\Xi^{(n)}(6758)\,L^{(n)}(8274)\,, \label{eq:nthorderHedin2}\\
    &\Sigma_c^{(n)}(12) = \nn \\
    &\qquad i v(1^{+}3)\,G^{(n-1)}(14)\,\Xi^{(n)}(4625)\,L^{(n)}(5363^+)\,, \label{eq:nthorderHedin3}\\
    &G^{(n)}(12) = G^{(0)}(12) \nn \\
    &\qquad \qquad \quad + G^{(0)}(13)\,\Delta\Sigma^{(n)}(34)\,G^{(n)}(42)\,,
    \label{eq:nthorderHedin4}
\end{align}
\noindent where $\Delta\Sigma^{(n)}(12) = \Sigma^{(n)}(12) - \Sigma^{(0)}(12)$, and
\begin{align}
    L^{(n)}_0(1234) = -i\,G^{(n-1)}(14)\,G^{(n-1)}(23)\,
\end{align}
\noindent is the non-interacting two-particle correlation function.

The superscript $(n)$ labels the cycle, not the order of a perturbative expansion. Provided that the iterative sequence converges and that $G^{(0)}$ is the Dyson solution for $\Sigma^{(0)}$, full self-consistency is recovered in the limit $n\rightarrow\infty$. At each iteration the total self-energy is written as
\begin{align} \label{eq:sigmapartition}
    \Sigma^{(n)}[G^{(n-1)}]
    \equiv&\,
    \Sigma_H[G^{(n-1)}]
    + \Sigma_x[G^{(n-1)}] \nn \\
    &\qquad \qquad \quad + \Sigma_c^{(n)}[G^{(n-1)}] .
\end{align}
Once $G^{(n)}$ has been obtained, all occurrences of $G^{(n-1)}$ in the next cycle are replaced by $G^{(n)}$. The sole exception is the reference Green's function $G^{(0)}$ in Eq.~\eqref{eq:nthorderHedin4}, which remains fixed in the Dyson equation.

In contrast to Hedin's equations, this formulation contains the BSE explicitly, as Eq.~\eqref{eq:nthorderHedin2}. It therefore gives direct access to neutral excitation energies, while Eq.~\eqref{eq:nthorderHedin3} gives the correlation part of the self-energy and hence access to charged excitations. The functional derivative in Eq.~\eqref{eq:nthorderHedin1} ensures that the kernel is consistent with the self-energy from the preceding iteration. The Dyson-like equations for $L$ and $G$ then resum, respectively, kernel insertions in the two-particle response and self-energy insertions in the one-particle Green's function.

\subsection{Truncation and starting points}
Full self-consistency in Eqs.~\eqref{eq:nthorderHedin1}--\eqref{eq:nthorderHedin4} is computationally not feasible. We therefore truncate the iterative sequence after the construction of $L^{(2)}$,
\begin{align}
    \{\Sigma^{(0)},G^{(0)}\}
    &\rightarrow \Xi^{(1)}
    \rightarrow L^{(1)}
    \rightarrow \Sigma^{(1)}
    \rightarrow G^{(1)} \nn \\
    &\rightarrow \Xi^{(2)}
    \rightarrow L^{(2)} \,.
\end{align}
We refer to this below as the second cycle.

All calculations are carried out in a fixed restricted Hartree--Fock (RHF) orbital basis. The distinction between the starting points is therefore not the choice of orbitals, but the QP energies and the corresponding self-energy approximation used to define $\Xi^{(1)}$.

\subsubsection{First cycle: HF starting point}

Our first choice is an HF starting point. In this case, $\Sigma^{(0)}$ contains the Hartree and bare exchange terms,
\begin{align}
    \Sigma^{(0)}_{\mathrm{HF}}(12)
    =& \underbrace{-i\,\delta(12)\,v(13)\,G^{(0)}(33^{+})}_{\Sigma_H(12)}
    \nn \\
    &\qquad \qquad \quad
    + \underbrace{i\,G^{(0)}(12)\,v(1^{+}2)}_{\Sigma_x(12)} \, .
    \label{eq:sigmaHF}
\end{align}
The corresponding non-interacting Green's function is built from HF orbitals and HF eigenvalues,
\begin{align}
    G^{(0)}_{\mathrm{HF}}(\mathbf{r},\mathbf{r}',\omega)
    =& \sum_{i}^{\mathrm{occ}}
    \frac{\varphi_i(\mathbf{r})\varphi_i(\mathbf{r}')}
    {\omega - \varepsilon_i^{\mathrm{HF}} - i\eta} \nn \\
    &\qquad + \sum_{a}^{\mathrm{virt}}
    \frac{\varphi_a(\mathbf{r})\varphi_a(\mathbf{r}')}
    {\omega - \varepsilon_a^{\mathrm{HF}} + i\eta} \, .
    \label{eq:g0HF}
\end{align}
Here $\varphi_p$ and $\varepsilon_p^{\mathrm{HF}}$ are the HF orbitals and eigenvalues, respectively. The indices $i,j,k,\dots$ label occupied states, $a,b,c,\dots$ label virtual states, and $\eta$ is a positive infinitesimal.

Taking the functional derivative $\Xi^{(1)} = i\,\delta\Sigma^{(0)}/\delta G^{(0)}$ then gives the TDHF kernel,
\begin{align} \label{eq:TDHFkernel}
    \Xi^{(1)}_{\mathrm{TDHF}}(1234)
    =& \delta(13)\delta(24)\,v(12)
    \nn \\
    &\qquad
    - \delta(14)\delta(32)\,v(13) \,.
\end{align}

\subsubsection{First cycle: $GW$ QP starting point}

As a second choice, we use a QP-corrected starting point. In this case we use the well-known $GW$ self-energy for $\Sigma^{(0)}$,
\begin{align}
    \Sigma^{(0)}_{GW}(12)
    =& \underbrace{-i\,\delta(12)\,v(13)\,G^{(0)}(33^{+})}_{\Sigma_H(12)}
    \nn \\
    &\;\;
    + \underbrace{i\,G^{(0)}(12)\,v(1^{+}2)}_{\Sigma_{x}(12)} \nn \\
    &  \quad + \underbrace{i\,G^{(0)}(12)\,\big(W(1^{+}2) - v(1^{+}2)\big)}_{\Sigma^{(0)}_{c}(12)}\, .
    \label{eq:sigmaGW}
\end{align}
The orbitals are still the HF orbitals, but the HF eigenvalues in $G^{(0)}$ are replaced by one-shot $G_0W_0$@HF QP energies. Thus,
\begin{align}
    G^{(0)}_{GW}(\mathbf{r},\mathbf{r}',\omega)
    =& \sum_{i}^{\mathrm{occ}}
    \frac{\varphi_i(\mathbf{r})\varphi_i(\mathbf{r}')}
    {\omega - \varepsilon_i^{GW} - i\eta} \nn \\
    &\qquad  + \sum_{a}^{\mathrm{virt}}
    \frac{\varphi_a(\mathbf{r})\varphi_a(\mathbf{r}')}
    {\omega - \varepsilon_a^{GW} + i\eta} \, .
    \label{eq:g0GW}
\end{align}
The functional derivative then gives
\begin{align} \label{eq:BSEkernel}
    \Xi^{(1)}_{\mathrm{BSE}}(1234)
    =& \delta(13)\delta(24)\,v(12)
    \nn \\
    &\qquad 
    - \delta(14)\delta(32)\,W_0(13) \, ,
\end{align}
where $W_0$ is the static screened interaction. In practice, $W_0$ is evaluated within the RPA in the same HF orbital basis at each cycle, using the QP energies of the updated Green's function. In the present case, these are those of $G^{(0)}_{GW}$. In deriving $\Xi^{(1)}_{\mathrm{BSE}}$ we neglect $\delta W/\delta G$, which is the standard approximation in BSE calculations.\cite{Hanke1979, Hanke1980, Strinati1988, Rohlfing2000,Onida2002,Blase2018}

Each choice of $\Sigma^{(0)}$ defines a distinct iterative sequence whose first cycle corresponds to a familiar method. The kernel $\Xi^{(1)}_{\mathrm{TDHF}}$ yields TDHF, whereas $\Xi^{(1)}_{\mathrm{BSE}}$ yields $GW$+BSE, which we denote as BSE$^{(1)}$@$GW$ from this point onward. The first-cycle self-energy and BSE/TDHF results correspond to those discussed in Ref.~\citenum{Forster2024}. Following that work, we denote $\Sigma^{(1)}$ of the HF and $GW$ sequences as $\Sigma^{\text{TDHF}}@L^{\text{TDHF}}$ and $\Sigma^{\text{BSE}}@L^{\text{BSE}}$, respectively. The label of $\Sigma$ specifies the kernel in the vertex of Eq.~\eqref{eq:nthorderHedin3} and the label of $L$ the kernel used to obtain $L^{(1)}$ from Eq.~\eqref{eq:nthorderHedin2}. In the present work, the resulting improved QP energies are then used to construct the second-cycle BSE, that is, to evaluate both $L^{(2)}_0$ and the screened interaction entering the kernel.

\subsubsection{Second cycle}
\label{sec:second-iteration}
The second cycle yields several methods, depending on the starting self-energy and on the interactions used in the two cycles. Table~\ref{tab:methods} summarizes them together with the first-cycle methods used for comparison.

The second-cycle construction is the same for the screened BSE and TDHF variants. The difference enters only through the expression used for the first-cycle kernel $\Xi^{(1)}$. Below we write the equations for the screened BSE case, with the TDHF variant obtained by replacing all occurrences of $W_0$ by the bare interaction $v$.

We denote the two second-cycle methods as BSE$^{(2)}$@HF and BSE$^{(2)}$@$GW$, where the superscript indicates the cycle at which the sequence is truncated and the label following @ specifies the starting self-energy $\Sigma^{(0)}$. When the interactions used in the two cycles differ, we indicate this by a subscript. BSE$^{(2)}_{Wv}$@$GW$ denotes a calculation in which $W_0$ enters $\Xi^{(1)}$ and the bare interaction $v$ enters $\Xi^{(2)}$. When no subscript is given, the same interaction is used in both cycles.

Using Eq.~\eqref{eq:sigmapartition}, the second-cycle kernel of Eq.~\eqref{eq:nthorderHedin1} separates into a Hartree--exchange part and a correlation part,
\begin{align} \label{eq:funcderiv}
    &\Xi^{(2)}(1234) \nn \\
    &= i\frac{\delta \Sigma^{(1)}(13)}{\delta G^{(1)}(42)} \nn \\
    &= \Xi_\text{TDHF}(1234) + i\frac{\delta \Sigma_c^{(1)}(13)}{\delta G^{(1)}(42)} \nn \\
    &= \Xi_\text{TDHF}(1234) + i\frac{\delta}{\delta G^{(1)}(42)} \nn \\
    &\quad\times \left\{i v(1^+7)\, G^{(1)}(18)\, \Xi^{(1)}(8635)\, L^{(1)}(5767^+)\right\} \,,
\end{align}
where $\Xi_\text{TDHF} = i\,\delta(\Sigma_H + \Sigma_x)/\delta G$ is the bare kernel of Eq.~\eqref{eq:TDHFkernel}, and the expression in braces is the first-cycle correlation self-energy of Eq.~\eqref{eq:nthorderHedin3}, evaluated with the updated propagator $G^{(1)}$. Equation~\eqref{eq:funcderiv} holds for both starting points, because $\Sigma_H$ and $\Sigma_x$ are the same functionals of $G$ in every cycle, and only the correlation part is replaced from one cycle to the next. In particular, the first-order part of the second-cycle kernel is the bare TDHF kernel also in the $GW$ sequence. The screening of the exchange term in $\Xi^{(1)}_\text{BSE}$, $-\delta(14)\delta(32)[W_0-v](13)$, originates from $\Sigma_c^{(0)}$, which $\Sigma_c^{(1)}$ replaces. In contrast to the first cycle, where the kernel is instantaneous, the correlation contribution to $\Xi^{(2)}$ is dynamical and contains both particle--hole (ph) and particle--particle (pp) propagation channels.

A direct evaluation of this derivative would require the derivative of the interacting response function $L^{(1)}$. Varying the Dyson-like equation for $L^{(1)}$ gives
\begin{align}
    \frac{\delta L^{(1)}}{\delta G^{(1)}}
        =& \,\big(1 + L^{(1)}\Xi^{(1)}\big)
          \big(\frac{\delta L_0^{(2)}}{\delta G^{(1)}}\big)
          \big(1 + \Xi^{(1)} L^{(1)}\big) \nn\\
          &+L^{(1)}\frac{\delta \Xi^{(1)}}{\delta G^{(1)}}L^{(1)}\,.
    \label{eq:dLdG}
\end{align}
Since $L_0$ depends explicitly on the Green's function, it is evaluated with the updated $G^{(1)}$, in accordance with the rule stated below Eq.~\eqref{eq:sigmapartition}. The derivative $\delta\Xi^{(1)}/\delta G^{(1)}$ in Eq.~\eqref{eq:dLdG} involves taking two functional derivatives of $\Sigma^{(0)}$. It therefore vanishes for $\Xi^{(1)}_{\mathrm{TDHF}}$, since $\Sigma^{(0)}_{\mathrm{HF}}$ is linear in $G^{(0)}$. For $\Xi^{(1)}_{\mathrm{BSE}}$, the same derivative vanishes once $\delta W/\delta G$ is neglected, consistent with the approximation employed in constructing $\Xi^{(1)}_{\mathrm{BSE}}$.

To keep the cost of the second cycle at that of second-order methods, we replace the interacting response function by its non-interacting counterpart evaluated with the updated Green's function,
\begin{equation}
    L^{(1)} \rightarrow L_0^{(2)} \,,
\end{equation}
already in Eq.~\eqref{eq:nthorderHedin3}, before taking the derivative. In Eq.~\eqref{eq:dLdG}, this amounts to setting the two dressed factors equal to the identity, leaving only
\begin{equation}
\frac{\delta L^{(1)}}{\delta G^{(1)}}
\rightarrow
\frac{\delta L_0^{(2)}}{\delta G^{(1)}}\,.
\end{equation}
For the HF sequence, and for BSE$^{(2)}_{Wv}$@$GW$, the truncated $\Sigma_c^{(1)}$ is then the second-order Green’s function (GF2) correlation self-energy, and $\Xi^{(2)}$ is the BSE@GF2 kernel. For the $GW$ sequence, the truncated $\Sigma_c^{(1)}$ also contains an exchange-type term in which one of the two interaction lines is $W_0$, as discussed below.

The replacement $L^{(1)} \rightarrow L_0^{(2)}$ also defines the power counting of the second-cycle kernel. Its correlation part contains all second-order diagrams that are irreducible in the ph channel of the BSE. $W_0$ enters only by dressing one of the two interaction lines in the exchange-type terms, which adds contributions of third and higher order in $v$. Unlike in BSE$^{(1)}$@$GW$, the first-order exchange term is not screened to all orders, as this would induce double counting. This is discussed in App.~\ref{app:doublecounting}. The resulting kernel has the same structure as the second-order BSE kernels found in  Refs.~\citenum{Rebolini2016,Zhang2013b,Monino2023},

\begin{strip}
\noindent\rule[5.4pt]{0.48\linewidth}{0.6pt}\rule[5.5pt]{0.4pt}{10pt}
\vspace{0.2cm}
\begin{align}\label{eq:second-cyclekernelbeforeeff}
    A_{ia,jb}^{(2)}(\omega)=& \sum_{kc} \bigg[ - \frac{\mathbf{2(ac|kb)(ic|jk)} - (ab|kc)(ic|W_0|jk) - (ac|kb)(kc|W_0|ij) + 2(ab|kc)(kc|ij)}{\omega-(\varepsilon_c-\varepsilon_k)+i\eta} \nn \\
    &+ \frac{\mathbf{2(bc|ka)(jc|ik)} - (ab|kc)(jc|W_0|ik) - (bc|ka)(kc|W_0|ij) + 2(ab|kc)(kc|ij)}{\omega+(\varepsilon_c-\varepsilon_k)-i\eta} \bigg]\nn\\ 
    &+ \sum_{kl} \frac{(kb|il)(ka|W_0|jl) \,\mathbf{-\, 2(kb|il)(la|jk)}}{\omega-(\varepsilon_k+\varepsilon_l)-i\eta} + \sum_{cd} \frac{\mathbf{2(cb|id)(ad|jc)} - (cb|id)(ca|W_0|jd)}{\omega-(\varepsilon_c+\varepsilon_d)+i\eta} \,.
\end{align}
\vspace{0.2cm}
\noindent\hfill\rule[-4pt]{0.4pt}{10pt}\rule[5.4pt]{0.48\linewidth}{0.6pt}
\end{strip}

The pp terms in this kernel involve sums of QP energies rather than energy differences. As a result, the dynamical kernel depends on the choice of chemical potential. To remove this dependence, we recast the dynamical kernel in the effective-kernel formulation of Ref.~\citenum{Rebolini2016} before taking the static limit. In this formulation, the ph and pp propagation contributions are contracted with the non-interacting response function to obtain an effective frequency-dependent kernel,
\begin{align}
    &\Xi^{(2)}_{\mathrm{eff}}(\omega)\nn\\
    &\;= L^{(2)}_0(\omega)^{-1} \int \frac{d\omega'}{2\pi} \frac{d\omega''}{2\pi}\, L^{(2)}_0(\omega',\omega)\nn \\
    & \;\;\;\quad\times \, \Xi^{(2,\mathrm{ph})}(\omega' - \omega'')L^{(2)}_0(\omega'',\omega)\, L^{(2)}_0(\omega)^{-1} \nonumber \\
    &\; \quad+ L^{(2)}_0(\omega)^{-1} \int \frac{d\omega'}{2\pi} \frac{d\omega''}{2\pi}\, L^{(2)}_0(\omega',\omega)\, \nn \\
    &\;\;\; \quad \times \Xi^{(2,\mathrm{pp})}(\omega' + \omega'')L^{(2)}_0(\omega'',\omega)\, L^{(2)}_0(\omega)^{-1} \,,
    \label{eq:effectivekernelsecondorder}
\end{align}
where $L^{(2)}_0(\omega',\omega)$ denotes the double Fourier transform defined in Ref.~\citenum{Rebolini2016}. The first term in Eq.~\eqref{eq:effectivekernelsecondorder} corresponds to the ph channel, while the second term corresponds to the pp channel. In this work, we employ the static approximation to the dynamical second-cycle kernel. Apart from the truncation of the iterative sequence, we thus make four approximations, namely (i) the quasiparticle form of $G^{(1)}$ (with fixed HF orbitals and no satellites), (ii) the replacement $L^{(1)} \rightarrow L_0^{(2)}$ in Eq.~\eqref{eq:nthorderHedin3} before differentiation, (iii) the neglect of $\delta W/\delta G$ in both cycles, and (iv) the static limit of the effective kernel. The resulting static kernel is combined with the instantaneous first-order TDHF kernel to construct the BSE Hamiltonian described below.

\subsubsection{Static working equations}

After the static approximation, the BSE can be solved by diagonalization in the ph basis,
\begin{equation}
    \begin{pmatrix} A & B \\ B & A \end{pmatrix}
    \begin{pmatrix} X \\ Y \end{pmatrix}
    = \begin{pmatrix} \Omega & 0 \\ 0 & -\Omega \end{pmatrix}
    \begin{pmatrix} X \\ Y \end{pmatrix}\,,
    \label{eq:bse-matrix}
\end{equation}
where $\Omega$ is a neutral excitation energy and $X$ and $Y$ are the corresponding excitation and de-excitation amplitudes. We report results for both the full BSE problem and the Tamm--Dancoff approximation (TDA).~\cite{Hirata1999} In the latter, the coupling block $B$ is set to zero, so that Eq.~\eqref{eq:bse-matrix} reduces to the following eigenvalue problem
\begin{equation}
    AX = \Omega X\,.
\end{equation}

The matrix blocks are decomposed into the first-order contribution and the static second-order correlation correction,
\begin{align}
    A &= A^{(1)} + A^{(2)} \,, \nn \\
    B &= B^{(1)} + B^{(2)} \,.
\end{align}

The first-cycle matrix elements follow from the standard TDHF and BSE kernels, Eqs.~\eqref{eq:TDHFkernel} and \eqref{eq:BSEkernel}. In the second cycle, the first-order part of the kernel is the bare TDHF kernel in both sequences (Eq.~\eqref{eq:funcderiv}), with matrix elements
\begin{align}
    A^{(1)}_{ia,jb} &= \delta_{ij}\delta_{ab}(\varepsilon_a - \varepsilon_i)
        + \mathbf{2(ia|jb)} - (ij|ab)\,, \nonumber\\
    B^{(1)}_{ia,jb} &= \mathbf{2(ia|bj)} - (ib|ja)\,,
    \label{eq:matrix-elements}
\end{align}
with two-electron integrals written in chemists' notation,
\begin{align}
    (pq|&K|rs) = \nn\\
    &\int d\mathbf{r} \int d\mathbf{r}'\,
        \varphi_p^*(\mathbf{r})\varphi_q(\mathbf{r})\,
        K(\mathbf{r},\mathbf{r}')\,
        \varphi_r^*(\mathbf{r}')\varphi_s(\mathbf{r}')\,,
    \label{eq:integrals}
\end{align}
and $(pq|rs) \equiv (pq|v|rs)$.

The second-cycle correction to the $A$ block follows from the resonant--resonant part of the BSE$^{(2)}$ Hamiltonian. Within the static approximation used here, we evaluate it in the limit $\omega\rightarrow 0$, which gives

\begin{strip}
\noindent\rule[5.4pt]{0.48\linewidth}{0.6pt}\rule[5.5pt]{0.4pt}{10pt}
\vspace{0.2cm}
\begin{align}\label{eq:second-cyclekernel}
    A_{ia,jb}^{(2)}=& \,\sum_{kc} \bigg[ \frac{\mathbf{2(ac|kb)(ic|jk)} - (ab|kc)(ic|W_0|jk) - (ac|kb)(kc|W_0|ij) + 2(ab|kc)(kc|ij)}{\varepsilon_b - \varepsilon_i + \varepsilon_c - \varepsilon_k} \nn \\
    &+ \frac{\mathbf{2(bc|ka)(jc|ik)} - (ab|kc)(jc|W_0|ik) - (bc|ka)(kc|W_0|ij) + 2(ab|kc)(kc|ij)}{\varepsilon_a - \varepsilon_j + \varepsilon_c - \varepsilon_k} \bigg]\nn\\ 
    &+ \sum_{kl} \frac{(kb|il)(ka|W_0|jl) \,\mathbf{- \,2(kb|il)(la|jk)}}{\varepsilon_a - \varepsilon_k + \varepsilon_b - \varepsilon_l} - \sum_{cd} \frac{\mathbf{2(cb|id)(ad|jc)} - (cb|id)(ca|W_0|jd)}{\varepsilon_c - \varepsilon_i + \varepsilon_d - \varepsilon_j} \,.
\end{align}
\vspace{0.2cm}
\noindent\hfill\rule[-4pt]{0.4pt}{10pt}\rule[5.4pt]{0.48\linewidth}{0.6pt}
\end{strip}
The coupling block $B^{(2)}_{ia,jb}$ is obtained from the resonant--anti-resonant matrix elements. As first observed in Ref.~\citenum{Rebolini2016}, the frequency arguments cancel in this sector, so that $B^{(2)}$ is frequency independent. $B^{(2)}_{ia,jb}$ follows from Eq.~\eqref{eq:second-cyclekernel} by interchanging the indices $j$ and $b$ in the numerators while keeping the denominators unchanged, after the replacement of the $W_0$-containing terms described below. The kernel has been spin-adapted, and the terms shown in bold cancel for triplet excitations.

Constructing the matrix elements in Eq.~\eqref{eq:second-cyclekernel} scales as $\mathcal{O}(N^5)$. Since we are interested only in the lowest neutral excitation energies, the resulting eigenvalue problem can be solved using the Davidson algorithm, retaining an overall $\mathcal{O}(N^5)$ scaling for the second-cycle calculation. The computational bottleneck is instead the first cycle, where obtaining the QP energies requires solving the full Casida problem. This step scales as $\mathcal{O}(N^6)$. Thus, the second cycle does not increase the formal scaling of the overall scheme.

A third starting point is provided by the GF2 self-energy. Taking $\Sigma^{(0)} = \Sigma_{\mathrm{GF2}}$ yields a first-cycle kernel of the same form as Eq.~\eqref{eq:second-cyclekernelbeforeeff} with $W_0$ replaced by $v$, but evaluated with GF2 QP energies. It is used here in the effective-kernel form of Eq.~\eqref{eq:second-cyclekernel}, rather than in the form conventionally used for BSE@GF2.\cite{Monino2023} The two are not identical, and the energies reported below are correspondingly not directly comparable with published BSE@GF2 values. We include it not as a new method but as a control. Since BSE$^{(1)}$@GF2 shares the kernel structure of BSE$^{(2)}$@HF and differs only in the QP energies inserted in the denominators, the pair isolates the effect of the QP spectrum from that of the kernel.

The kernel shared by BSE$^{(1)}$@GF2 and BSE$^{(2)}$@HF also relates both methods to second-order algebraic diagrammatic construction [ADC(2)],~\cite{Schirmer1982} which we include as a reference in Sec.~\ref{sec:results}. Monino and Loos showed that, within the TDA, the dynamical BSE@GF2 problem can be upfolded into a linear eigenvalue problem in the space of single (1h1p) and double (2h2p) excitations, whose symmetrized form has the same 2h2p block and 1h1p--2h2p coupling as ADC(2).\cite{Monino2023} Upfolding to the double-excitation space generates both the full frequency dependence of the kernel and the GF2 renormalization of the orbital energies.

\subsubsection{Symmetrization of the exchange-type term}
For the $GW$ starting point, inserting Eq.~\eqref{eq:BSEkernel} into Eq.~\eqref{eq:nthorderHedin3} splits the correlation self-energy $\Sigma_c^{(1)}$ into two terms. The $v$ part of $\Xi^{(1)}_{\mathrm{BSE}}$ gives a ring term with two bare interactions. The $W_0$ part gives an exchange-type term,
\begin{equation}
    \Sigma_c^{(1b)}(12) = -v(1^+3)G(14)G(43^+)G(32)W_0(42) \,,
    \label{eq:sigmac-1b}
\end{equation}
in which $v$ is attached to point 1 and $W_0$ to point 2, the coordinates of the final self-energy. All $W_0$-containing terms of $A^{(2)}$ and $B^{(2)}$ derive from Eq.~\eqref{eq:sigmac-1b}, with $W_0$ on the integral that originates from point 2. As a result, the matrices are not symmetric. For example, the $kl$ term $(kb|il)(ka|W_0|jl)$ of $A^{(2)}_{ia,jb}$ appears as $(ka|jl)(kb|W_0|il)$ in $A^{(2)}_{jb,ia}$, and the two coincide only for $W_0=v$.

\begin{figure}
  \includegraphics[width=\columnwidth]{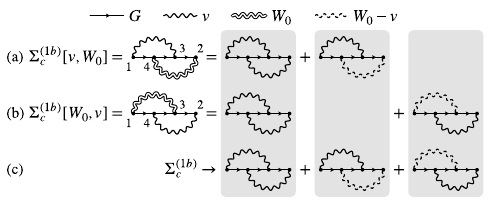}
  \caption{Exchange-type term $\Sigma_c^{(1b)}$ of the correlation self-energy for the $GW$ starting point, and its symmetrization. In each diagram, the upper interaction line ($X$ in $\Sigma_c^{(1b)}[X,Y]$) is attached to the external point 1 and the lower one ($Y$) to the external point 2. (a) The equation of motion with respect to the first argument of $G$ gives $\Sigma_c^{(1b)}[v,W_0]$. (b) The equation of motion with respect to the second argument gives the mirror form $\Sigma_c^{(1b)}[W_0,v]$, which for real orbitals is the transpose of (a). Both contain the bare term $\Sigma_c^{(1b)}[v,v]$ but differ in their screening corrections. (c) Eq.~\eqref{eq:2sosex} adds (a) and (b) and subtracts the bare term once.}
  \label{fig:sosex}
\end{figure}

Equation~\eqref{eq:nthorderHedin3} attaches $v$ to point 1 because it follows from the equation of motion of $G$ with respect to its first argument. The equation of motion with respect to the second argument gives a mirror form (Figure~\ref{fig:sosex}), in which $v$ is attached to point 2 (see, for instance, Ref.~\citenum{Strinati1988}). The two forms give the same self-energy for the exact kernel, but not for an approximate kernel such as $\Xi^{(1)}_{\mathrm{BSE}}$. For the HF starting point both interactions are bare, and the two forms coincide.

To make the difference explicit, we write $\Sigma_c^{(1b)}[X,Y]$ for the term of Eq.~\eqref{eq:sigmac-1b} with the interaction $X$ attached to point 1 and $Y$ to point 2, so that the two forms are $\Sigma_c^{(1b)}[v,W_0]$ and $\Sigma_c^{(1b)}[W_0,v]$. When $W_0$ is expanded in powers of $v$, both forms start with the bare second-order exchange term $\Sigma_c^{(1b)}[v,v]$, whereas their higher-order diagrams are distinct and each appears once in the exact self-energy. We therefore include both forms and count the common bare term once,
\begin{equation}
    \Sigma_c^{(1b)} \rightarrow \Sigma_c^{(1b)}[v,W_0] + \Sigma_c^{(1b)}[W_0,v] - \Sigma_c^{(1b)}[v,v] \,.
    \label{eq:2sosex}
\end{equation}
The right-hand side is the statically screened analogue of the 2SOSEX self-energy of Ref.~\citenum{Bruneval2024}, whereas $\Sigma_c^{(1b)}[v,W_0]$ alone corresponds to the SOSEX self-energy of Ref.~\citenum{Ren2015}. In the matrix elements of $A^{(2)}$ and $B^{(2)}$, every term that contains $W_0$ is replaced accordingly, for example
\begin{align}
    (ab|kc)(ic|W_0|jk) \rightarrow&\; (ab|kc)(ic|W_0|jk) \nn \\
    &\qquad + (ab|W_0|kc)(ic|jk) \nn\\
    &\qquad \qquad - (ab|kc)(ic|jk) \,,
    \label{eq:w0union}
\end{align}
with the denominator unchanged. For real orbitals, the two mirror forms are transposes of each other, so the resulting $A^{(2)}$ and $B^{(2)}$ are symmetric.

\subsubsection{Screened exchange classification}
Equation~\eqref{eq:sigmapartition} prescribes that the correlation self-energy of cycle $n$ replaces that of cycle $n - 1$, while the Hartree and exchange terms are kept with the bare interaction. For the $GW$ starting point, this requires the partition of Eq.~\eqref{eq:sigmaGW}, in which $iG(W-v)$ is assigned to $\Sigma_c^{(0)}$. The alternative partition
\begin{align}
    \Sigma^{(0)}_{GW}(12)
    =& \underbrace{-i\,\delta(12)\,v(13)\,G^{(0)}(33^{+})}_{\Sigma_H(12)}
    \nn \\
    &\qquad \quad
    + \underbrace{i\,G^{(0)}(12)\,W(1^{+}2)}_{\Sigma_{x}(12)} \,,
    \label{eq:sigmaGWex}
\end{align}
assigns all of $iGW$ to the exchange term. Because the exchange term is kept in every cycle, the first-order blocks $A^{(1)}$ and $B^{(1)}$ of the second cycle then contain the screened terms $-(ij|W_0|ab)$ and $-(ib|W_0|ja)$ instead of the bare ones of Eq.~\eqref{eq:matrix-elements}. The second-order part of $-(ij|W_0|ab)$ contains the same integrals, with the same sign, as terms already present in $A^{(2)}$ and is therefore counted twice (App.~\ref{app:doublecounting}). Using this partition raises the valence excitation energies by more than 1~eV on average, five to seven times more than the Rydberg excitation energies, and increases the mean absolute error three- to fourfold (Supporting Information).

\subsection{Dynamical perturbation}

Several strategies have been proposed to go beyond the static approximation to the BSE kernel.~\cite{Authier2021,Bintrim2022} In this work, we include the frequency dependence of the effective kernel perturbatively, through a renormalized first-order correction to the static BSE$^{(2)}$ excitation energies, which we denote dBSE$^{(2)}$. This correction accounts, in an approximate way, for additional relaxation effects associated with higher excitations. It follows the first-order perturbative treatment of dynamical screening of Rohlfing and co-workers,~\cite{Rohlfing2000,Ma2009,Ma2010,Baumeier2012a} in the renormalized form that Loos and Blase applied to the dynamically screened kernel beyond the plasmon-pole approximation,~\cite{Loos2020b} and that has also been used for second-order kernels.~\cite{Zhang2013b,Rebolini2016,Monino2023}

The dBSE$^{(2)}$ excitation energy of state $m$ is written as
\begin{equation}
    \Omega^{\mathrm{dBSE^{(2)}}}_{m}
    =
    \Omega^{\mathrm{BSE^{(2)}}}_{m}
    +
    \zeta_m\,\delta\Omega_m
    \,,
    \label{eq:dbse}
\end{equation}
where $\Omega^{\mathrm{BSE^{(2)}}}_{m}$ is the static BSE excitation energy, $\delta\Omega_m$ is the first-order dynamical correction, and $\zeta_m$ is a renormalization factor, both defined below.

We define the dynamical part of the effective kernel as the difference between the frequency-dependent kernel and its static limit,
\begin{equation}
    \Delta\Xi_{\mathrm{eff},(ia,jb)}(\omega)
    =
    \Xi_{\mathrm{eff},(ia,jb)}(\omega)
    -
    \Xi_{\mathrm{eff},(ia,jb)}(0) \, .
    \label{eq:dxi}
\end{equation}
In the particle--hole representation this defines dynamical corrections $\Delta A(\omega)$ and $\Delta B$ to the BSE matrix blocks.

Since Eq.~\eqref{eq:bse-matrix} is not a symmetric eigenvalue problem, its left and right eigenvectors differ. We write it as $H R_m = \Omega_m R_m$, where $H=\sigma M$. Here, $M$ is the symmetric matrix on the left-hand side of Eq.~\eqref{eq:bse-matrix}, and $\sigma=\mathrm{diag}(1,-1)$. The right eigenvector is then given by $R_m=(X_m, Y_m)$, while the left eigenvector is $L_m = \sigma R_m = (X_m,\, -Y_m)$. First-order perturbation theory in the dynamical part of the kernel then gives
\begin{align}
    &\delta \Omega_m(\omega) \nn \\
    &\,=
    \begin{pmatrix}
        X_m^T & -Y_m^T
    \end{pmatrix}
    \begin{pmatrix}
        \Delta A(+\omega) & \Delta B \\
        -\Delta B & -\Delta A(-\omega)
    \end{pmatrix}
    \begin{pmatrix}
        X_m \\
        Y_m
    \end{pmatrix}
    \nn \\
    &\,=
    X_m^{T}\Delta A(+\omega)X_m
    +
    Y_m^{T}\Delta A(-\omega)Y_m
    \nn \\
    &\qquad \qquad \qquad \quad +
    X_m^{T}\Delta B\,Y_m
    +
    Y_m^{T}\Delta B\,X_m \, ,
    \label{eq:dbse-full-corr}
\end{align}
where $X_m$ and $Y_m$ are the static BSE eigenvectors of state $m$. The correction entering Eq.~\eqref{eq:dbse} is evaluated at the static excitation energy,
\begin{equation}
    \delta\Omega_m
    =
    \delta\Omega_m(\Omega_m^{\mathrm{BSE^{(2)}}}) \, .
    \label{eq:dbse-corr}
\end{equation}

For the effective kernels used here, the coupling block of the second-cycle kernel is frequency independent, so that $\Delta B = 0$. The $B$ terms in Eq.~\eqref{eq:dbse-full-corr} therefore vanish, and the correction reduces to
\begin{equation}
    \delta\Omega_m(\omega)
    =
    X_m^{T}\Delta A(+\omega)X_m
    +
    Y_m^{T}\Delta A(-\omega)Y_m \, .
    \label{eq:dbse-ablock-corr}
\end{equation}
In the TDA, $Y_m=0$ and only the first term remains.

The renormalization factor follows from linearizing the condition $\omega = \Omega_m^{\mathrm{BSE^{(2)}}} + \delta\Omega_m(\omega)$ around $\omega=\Omega_m^{\mathrm{BSE^{(2)}}}$,
\begin{equation}
    \zeta_m
    =
    \left[
    1
    -
    \left.
    \frac{\partial \delta\Omega_m(\omega)}
    {\partial \omega}
    \right|_{\omega=\Omega_m^{\mathrm{BSE^{(2)}}}}
    \right]^{-1} ,
    \label{eq:zeta}
\end{equation}
with
\begin{align}
    \frac{\partial\,\delta\Omega_m(\omega)}{\partial\omega}
    &= \,X_m^T \left.\frac{\partial A^{(2)}(\omega')}{\partial\omega'}\right|_{\omega'=\omega} X_m \nn \\
    &\qquad\quad - Y_m^T \left.\frac{\partial A^{(2)}(\omega')}{\partial\omega'}\right|_{\omega'=-\omega} Y_m \,.
    \label{eq:dzeta}
\end{align}
Equation~\eqref{eq:dbse} is thus a single Newton step toward the solution of the frequency-dependent eigenvalue problem, and agrees with it to first order in the dynamical part of the kernel.

The factor $\zeta_m$ also serves as a diagnostic of the perturbative treatment. Unlike the QP renormalization factor in $GW$, it is not a spectral weight and is not restricted to the interval $(0,1]$. Values close to one indicate that the kernel varies weakly with frequency around $\Omega_m^{\mathrm{BSE^{(2)}}}$, and that the first-order correction is reliable. A large deviation of $\zeta_m$ from one signals a near-pole situation, and a value $\zeta_m \le 0$ means that the linearization has broken down. %
\section{Computational details}
\label{sec:CompDetails}

All calculations were performed with the BAND~\cite{TeVelde1991} engine in a modified development version of the Amsterdam Modeling Suite (AMS2025.206),~\cite{Baerends2025} while the implementations of the second-cycle kernel and dynamical correction were independently verified against QuAcK.~\cite{Loos2019QuAcK} We use the spin-restricted formalism throughout, as all systems considered have closed-shell singlet ground states, and retain the default AMS settings unless stated otherwise. The preceding mean-field calculations were performed at the HF level. The implementation supports both Slater-type orbitals~\cite{vanLenthe2003} and Gaussian-type orbitals~\cite{Dunning1989,Kendall1992} obtained from the Basis Set Exchange.~\cite{Pritchard2019} In all calculations reported here, we use Gaussian-type orbitals and set the imaginary broadening $\eta$ equal to 0.001 Ha in the self-energy, and to 0 in the kernel.

Our test set comprises the seven molecules considered in Ref.~\citenum{Monino2023}. We use the geometries from that work, which are reproduced in the Supporting Information, and compute singlet and triplet excitation energies in the aug-cc-pVTZ basis. Theoretical best estimates (TBEs) and ADC(2) reference values in the aug-cc-pVTZ basis are taken from version 1.2 of the QUEST database.~\cite{Loos2018b, Veril2021,Loos2025QUEST} These TBEs are estimated to lie within $0.05$~eV of the corresponding FCI/aug-cc-pVTZ values.~\cite{Loos2025QUEST}

All statistical measures are evaluated with respect to the TBEs. We report the mean absolute error (MAE), mean signed error (MSE), root-mean-square error (RMSE), and maximum absolute error (Max). Given the modest size of the test set, small differences in MAE between methods should be regarded as indicative rather than significant.

Calculated excitations are assigned to reference states on the basis of their symmetry and character. Energy ordering is used only to distinguish states within a given symmetry block. Symmetry labels are obtained from the transition vectors. The state character and valence/Rydberg classification reported in the reference database are used as additional constraints on the hole orbital and, for valence states, on the virtual orbital carrying valence character. These constraints are of course not applied when the reference character is listed as undetermined. %
\section{Results}
\label{sec:results}

\subsection{Benchmark strategy}
\label{sec:benchmark}

\begin{table}[hbt!]
\footnotesize
\caption{Methods compared in this work. The QP column specifies the QP energies entering the BSE kernel in the final cycle. The first- and second-cycle columns indicate the interaction used to construct the corresponding kernels. $\Sigma^{\text{TDHF}}$ and $\Sigma^{\text{BSE}}$ abbreviate $\Sigma^{\text{TDHF}}@L^{\text{TDHF}}$ and $\Sigma^{\text{BSE}}@L^{\text{BSE}}$.}
\begin{tabular}{@{}lllcc@{}}
\hline
Method & $\Sigma^{(0)}$ & QP & 1st cycle & 2nd cycle \\
\hline
TDHF                       & $\Sigma_{\mathrm{HF}}$  & HF     & $v$   & --  \\
BSE$^{(1)}$@$GW$           & $\Sigma_{GW}$      & $G_0W_0$ & $W_0$ & --  \\
BSE$^{(1)}$@GF2            & $\Sigma_{\mathrm{GF2}}$ & GF2    & $v$   & --  \\
BSE$^{(2)}$@HF             & $\Sigma_{\mathrm{HF}}$  & $\Sigma^{\text{TDHF}}$     & $v$   & $v$ \\
BSE$^{(2)}$@$GW$           & $\Sigma_{GW}$      &  $\Sigma^{\text{BSE}}$  & $W_0$ & $W_0$ \\
BSE$^{(2)}_{Wv}$@$GW$    & $\Sigma_{GW}$      &  $\Sigma^{\text{BSE}}$  & $W_0$ & $v$ \\
\hline\label{tab:methods}
\end{tabular}
\end{table}

The methods included in the benchmark are summarized in Table~\ref{tab:methods}. The BSE$^{(1)}$@$GW$ results were generated in the present work. BSE$^{(1)}$@GF2 uses the effective kernel of Eq.~\eqref{eq:second-cyclekernel} for $W_0 = v$, evaluated with GF2 QP energies. BSE$^{(2)}$@HF uses the same kernel, evaluated with the $G^{(1)}$ energies of the HF sequence, which are obtained from $\Sigma^{\text{TDHF}}@L^{\text{TDHF}}$. BSE$^{(2)}$@$GW$ is evaluated with the $G^{(1)}$ energies of the $GW$ sequence, obtained from $\Sigma^{\text{BSE}}@L^{\text{BSE}}$. Here, $W_0$ enters the first-cycle kernel and the exchange-type terms of the second-order part of the second-cycle kernel. BSE$^{(2)}_{Wv}$@$GW$, introduced in Sec.~\ref{sec:second-iteration}, is included as a diagnostic. Since it differs from BSE$^{(2)}$@$GW$ only in the interaction entering the second-cycle kernel, the comparison isolates the effect of reusing $W_0$ in the second cycle.

Table~\ref{tab:singlet_full} lists the singlet and Table~\ref{tab:triplet_full} the triplet excitation energies obtained from the full, non-TDA eigenvalue problem in Eq.~\eqref{eq:bse-matrix}. For comparison, both tables also include ADC(2) results. Dynamically corrected values are given in parentheses for every method whose kernel contains the effective second-cycle form in Eq.~\eqref{eq:second-cyclekernel}. For BSE$^{(1)}$@$GW$ we do not consider such a dynamical correction. Results obtained within the TDA are reported separately in Tables~\ref{tab:singlet_tda} and~\ref{tab:triplet_tda}.

\providecommand{\stkhdr}[2]{\shortstack{#1\\[0.2ex]{\scriptsize$#2$}}}
\providecommand{\stkone}[1]{\shortstack{#1\\[0.2ex]{\scriptsize\phantom{$W^{(1)}$}}}}
 
\providecommand{\subGW}{W^{(1)}v^{(2)}_1W^{(2)}_2}
\providecommand{\subdc}{W^{(1)}W^{(2)}_1W^{(2)}_2}
\providecommand{\subv} {W^{(1)}W^{(2)}_1v^{(2)}_2}
\providecommand{\subWv}{W^{(1)}v^{(2)}_1v^{(2)}_2}
 
\providecommand{\mBSEGW}   {BSE$^{(2)}_{\subGW}$@$GW$}
\providecommand{\mBSEdc}   {BSE$^{(2)}_{\subdc}$@$GW$}
\providecommand{\mBSEv}    {BSE$^{(2)}_{\subv}$@$GW$}
\providecommand{\mBSEWv}   {BSE$^{(2)}_{\subWv}$@$GW$}
\providecommand{\mBSEHF}   {BSE$^{(2)}$@HF}
\providecommand{\mBSEGFtwo}{BSE$^{(1)}$@GF2}
\providecommand{\mBSEoneGW}{BSE$^{(1)}$@$GW$}
 
\providecommand{\hBSEGW}      {\stkhdr{BSE$^{(2)}$@$GW$}{\subGW}}
\providecommand{\hBSEGWmain}   {\stkone{BSE$^{(2)}$@$GW$}}
\providecommand{\hBSEGWtda}   {\stkhdr{BSE$^{(2)}$@$GW\backslash$TDA}{\subGW}}
\providecommand{\hBSEdc}      {\stkhdr{BSE$^{(2)}$@$GW$}{\subdc}}
\providecommand{\hBSEdctda}   {\stkhdr{BSE$^{(2)}$@$GW\backslash$TDA}{\subdc}}
\providecommand{\hBSEv}       {\stkhdr{BSE$^{(2)}$@$GW$}{\subv}}
\providecommand{\hBSEvtda}    {\stkhdr{BSE$^{(2)}$@$GW\backslash$TDA}{\subv}}
\providecommand{\hBSEWv}      {\stkhdr{BSE$^{(2)}$@$GW$}{\subWv}}
\providecommand{\hBSEWvmain}   {\stkone{BSE$^{(2)}_{Wv}$@$GW$}}
\providecommand{\hBSEHF}      {\stkone{BSE$^{(2)}$@HF}}
\providecommand{\hBSEHFtda}   {\stkone{BSE$^{(2)}$@HF$\backslash$TDA}}
\providecommand{\hBSEGFtwo}   {\stkone{BSE$^{(1)}$@GF2}}
\providecommand{\hBSEGFtwotda}{\stkone{BSE$^{(1)}$@GF2$\backslash$TDA}}
\providecommand{\hBSEoneGW}   {\stkone{BSE$^{(1)}$@$GW$}}
\providecommand{\hCCSD}       {\stkone{CCSD}}
\providecommand{\hTBE}        {\stkone{TBE}}
 
\begin{table*}[hbt!]
\setlength{\aboverulesep}{0.25ex}\setlength{\belowrulesep}{0.4ex}
\providecommand{\stt}[1]{$\vphantom{X}$\smash{#1}}
\providecommand{\hADCtwo}{\stkone{ADC(2)}}
\renewcommand{\stkone}[1]{#1}
\centering
\scriptsize
\setlength{\tabcolsep}{3pt}
\renewcommand{\arraystretch}{0.9}
\caption{Singlet excitation energies (in eV, aug-cc-pVTZ), dynamically corrected (dBSE) values in parentheses. Statistics are errors with respect to the TBE, for the whole set and separately for valence (Val.) and Rydberg (Ryd.) states. The two states of mixed valence--Rydberg character are included in the overall statistics but in neither subset.}
\label{tab:singlet_full}
\begin{tabular*}{\textwidth}{@{\extracolsep{\fill}}lll c c c c c c c}
\toprule
\stkone{Mol.} & \stkone{State} & \stkone{Nat.} & \hADCtwo & \hBSEGFtwo & \hBSEoneGW & \hBSEGWmain & \hBSEHF & \hBSEWvmain & \hTBE \\
\midrule
HCl & \stt{$^1\Pi$} & Val. & 7.97 & 7.82 (7.95) & 8.28 & 8.02 (8.12) & 8.17 (8.29) & 8.13 (8.25) & 7.84 \\
H$_2$O & \stt{$^1B_1$} & Ryd. & 7.18 & 6.85 (6.96) & 8.08 & 7.77 (7.86) & 8.07 (8.19) & 7.79 (7.90) & 7.63 \\
 & \stt{$^1A_2$} & Ryd. & 8.84 & 8.49 (8.58) & 9.76 & 9.47 (9.56) & 9.71 (9.82) & 9.42 (9.52) & 9.50 \\
 & \stt{$^1A_1$} & Ryd. & 9.52 & 9.21 (9.33) & 10.41 & 10.07 (10.15) & 10.32 (10.44) & 10.12 (10.24) & 9.99 \\
N$_2$ & \stt{$^1\Pi_g$} & Val. & 9.48 & 8.84 (9.22) & 10.43 & 9.65 (9.96) & 10.01 (10.42) & 9.81 (10.22) & 9.32 \\
 & \stt{$^1\Sigma_u^-$} & Val. & 10.26 & 10.07 (10.54) & 10.13 & 10.14 (10.66) & 9.78 (10.21) & 10.06 (10.51) & 9.88 \\
 & \stt{$^1\Delta_u$} & Val. & 10.79 & 10.50 (10.95) & 10.78 & 10.37 (10.82) & 10.22 (10.64) & 10.49 (10.93) & 10.29 \\
 & \stt{$^1\Sigma_g^+$} & Ryd. & 12.99 & 12.52 (12.57) & 13.90 & 13.39 (13.40) & 13.72 (13.77) & 13.46 (13.52) & 12.97 \\
 & \stt{$^1\Pi_u$} & Mix. & 13.33 & 12.87 (12.92) & 14.18 & 13.62 (13.56) & 14.20 (14.28) & 13.79 (13.86) & 13.09 \\
 & \stt{$^1\Sigma_u^+$} & Ryd. & 13.07 & 12.70 (12.81) & 13.98 & 13.54 (13.74) & 13.72 (13.95) & 13.57 (13.72) & 13.09 \\
 & \stt{$^1\Pi_u$} & Mix. & 13.99 & 13.05 (13.51) & 15.07 & 14.21 (14.41) & 14.75 (15.20) & 14.27 (14.69) & 13.42 \\
 & \stt{$^1\Pi_u$} & Ryd. & 14.55 & 14.26 (14.37) & 14.32 & 14.07 (14.28) & 13.88 (13.97) & 14.23 (14.35) & 14.01 \\
CO & \stt{$^1\Pi$} & Val. & 8.69 & 8.41 (8.73) & 9.54 & 8.91 (9.20) & 9.13 (9.46) & 9.02 (9.34) & 8.48 \\
 & \stt{$^1\Sigma^-$} & Val. & 10.03 & 9.56 (9.94) & 10.25 & 9.98 (10.30) & 10.09 (10.47) & 10.07 (10.45) & 9.93 \\
 & \stt{$^1\Delta$} & Val. & 10.30 & 9.79 (10.12) & 10.72 & 9.99 (10.20) & 10.33 (10.66) & 10.31 (10.64) & 10.07 \\
 & \stt{$^1\Sigma^+$} & Ryd. & 11.32 & 11.08 (11.09) & 11.99 & 11.48 (11.43) & 11.74 (11.76) & 11.61 (11.62) & 10.96 \\
 & \stt{$^1\Sigma^+$} & Ryd. & 11.83 & 11.58 (11.62) & 12.36 & 12.08 (12.12) & 12.24 (12.28) & 12.09 (12.14) & 11.52 \\
 & \stt{$^1\Pi$} & Ryd. & 12.03 & 11.50 (11.56) & 12.32 & 12.08 (12.15) & 12.19 (12.26) & 12.07 (12.14) & 11.72 \\
C$_2$H$_2$ & \stt{$^1\Sigma_u^-$} & Val. & 7.24 & 7.12 (7.45) & 7.38 & 7.47 (7.82) & 7.23 (7.54) & 7.44 (7.76) & 7.09 \\
 & \stt{$^1\Delta_u$} & Val. & 7.56 & 7.39 (7.71) & 7.75 & 7.63 (7.96) & 7.49 (7.80) & 7.70 (8.02) & 7.43 \\
C$_2$H$_4$ & \stt{$^1B_{3u}$} & Ryd. & 7.34 & 7.20 (7.25) & 7.59 & 7.54 (7.58) & 7.41 (7.46) & 7.57 (7.62) & 7.37 \\
 & \stt{$^1B_{1u}$} & Val. & 7.91 & 7.69 (7.99) & 8.17 & 8.03 (8.32) & 8.02 (8.28) & 8.19 (8.47) & 7.90 \\
 & \stt{$^1B_{1g}$} & Ryd. & 7.99 & 7.85 (7.91) & 8.23 & 8.21 (8.27) & 8.06 (8.12) & 8.23 (8.28) & 8.04 \\
CH$_2$O & \stt{$^1A_2$} & Val. & 3.92 & 3.96 (4.20) & 5.06 & 4.89 (5.12) & 5.10 (5.39) & 5.05 (5.34) & 3.97 \\
 & \stt{$^1B_2$} & Ryd. & 6.51 & 6.36 (6.34) & 7.84 & 7.31 (7.27) & 7.49 (7.48) & 7.39 (7.38) & 7.22 \\
 & \stt{$^1B_2$} & Ryd. & 7.53 & 7.35 (7.40) & 8.75 & 8.33 (8.38) & 8.47 (8.53) & 8.38 (8.44) & 8.12 \\
 & \stt{$^1A_1$} & Ryd. & 7.47 & 7.36 (7.35) & 8.83 & 8.34 (8.32) & 8.49 (8.48) & 8.39 (8.38) & 8.22 \\
 & \stt{$^1A_2$} & Ryd. & 7.99 & 7.30 (7.32) & 8.85 & 8.39 (8.39) & 8.53 (8.54) & 8.46 (8.48) & 8.66 \\
 & \stt{$^1B_1$} & Val. & 9.17 & 8.69 (9.26) & 10.20 & 9.70 (10.17) & 9.96 (10.56) & 9.84 (10.44) & 9.19 \\
 & \stt{$^1A_1$} & Val. & 9.46 & 9.13 (9.41) & 10.06 & 9.51 (9.85) & 9.73 (10.12) & 9.81 (10.20) & 9.43 \\
\midrule
MAE &  &  & 0.30 & 0.39 (0.36) & 0.63 & 0.29 (0.45) & 0.43 (0.61) & 0.37 (0.56) &  \\
MSE &  &  & 0.00 & $-$0.33 ($-$0.13) & +0.63 & +0.26 (+0.43) & +0.40 (+0.60) & +0.35 (+0.55) &  \\
RMSE &  &  & 0.38 & 0.51 (0.49) & 0.72 & 0.36 (0.53) & 0.55 (0.74) & 0.43 (0.65) &  \\
Max &  &  & 0.75 & 1.36 (1.34) & 1.65 & 0.92 (1.15) & 1.33 (1.78) & 1.08 (1.37) &  \\
\midrule
MAE &  & Val. & 0.16 & 0.21 (0.22) & 0.61 & 0.28 (0.59) & 0.37 (0.69) & 0.39 (0.75) &  \\
MSE &  & Val. & +0.15 & $-$0.14 (+0.20) & +0.61 & +0.27 (+0.59) & +0.34 (+0.69) & +0.39 (+0.75) &  \\
RMSE &  & Val. & 0.21 & 0.27 (0.31) & 0.69 & 0.36 (0.65) & 0.49 (0.79) & 0.46 (0.80) &  \\
Max &  & Val. & 0.50 & 0.50 (0.66) & 1.11 & 0.92 (1.15) & 1.13 (1.42) & 1.08 (1.37) &  \\
\midrule
MAE &  & Ryd. & 0.40 & 0.55 (0.52) & 0.55 & 0.24 (0.29) & 0.37 (0.42) & 0.29 (0.34) &  \\
MSE &  & Ryd. & $-$0.19 & $-$0.49 ($-$0.44) & +0.55 & +0.20 (+0.26) & +0.33 (+0.40) & +0.25 (+0.31) &  \\
RMSE &  & Ryd. & 0.47 & 0.67 (0.63) & 0.61 & 0.30 (0.34) & 0.44 (0.51) & 0.33 (0.39) &  \\
Max &  & Ryd. & 0.75 & 1.36 (1.34) & 1.03 & 0.56 (0.65) & 0.78 (0.86) & 0.65 (0.66) &  \\
\bottomrule
\end{tabular*}
\end{table*}
 
\begin{table*}[hbt!]
\setlength{\aboverulesep}{0.25ex}\setlength{\belowrulesep}{0.4ex}
\providecommand{\stt}[1]{$\vphantom{X}$\smash{#1}}
\providecommand{\hADCtwo}{\stkone{ADC(2)}}
\renewcommand{\stkone}[1]{#1}
\centering
\scriptsize
\setlength{\tabcolsep}{3pt}
\renewcommand{\arraystretch}{0.9}
\caption{Triplet excitation energies (in eV, aug-cc-pVTZ), dynamically corrected (dBSE) values in parentheses. Statistics are errors with respect to the TBE, for the whole set and separately for valence (Val.) and Rydberg (Ryd.) states.}
\label{tab:triplet_full}
\begin{tabular*}{\textwidth}{@{\extracolsep{\fill}}lll c c c c c c c}
\toprule
\stkone{Mol.} & \stkone{State} & \stkone{Nat.} & \hADCtwo & \hBSEGFtwo & \hBSEoneGW & \hBSEGWmain & \hBSEHF & \hBSEWvmain & \hTBE \\
\midrule
H$_2$O & \stt{$^3B_1$} & Ryd. & 6.86 & 6.54 (6.70) & 7.61 & 7.48 (7.63) & 7.75 (7.92) & 7.47 (7.64) & 7.25 \\
 & \stt{$^3A_2$} & Ryd. & 8.72 & 8.37 (8.50) & 9.56 & 9.37 (9.49) & 9.58 (9.73) & 9.29 (9.44) & 9.24 \\
 & \stt{$^3A_1$} & Ryd. & 9.15 & 8.79 (9.00) & 9.79 & 9.65 (9.82) & 9.88 (10.09) & 9.69 (9.90) & 9.54 \\
N$_2$ & \stt{$^3\Sigma_u^+$} & Val. & 8.15 & 8.05 (8.46) & 8.07 & 8.28 (8.72) & 7.78 (8.16) & 8.04 (8.44) & 7.70 \\
 & \stt{$^3\Pi_g$} & Val. & 8.20 & 7.55 (8.00) & 8.69 & 8.43 (8.86) & 8.69 (9.17) & 8.50 (8.98) & 8.02 \\
 & \stt{$^3\Delta_u$} & Val. & 9.25 & 9.14 (9.60) & 9.08 & 9.20 (9.68) & 8.85 (9.28) & 9.12 (9.57) & 8.87 \\
 & \stt{$^3\Sigma_u^-$} & Val. & 10.23 & 10.04 (10.47) & 10.13 & 10.15 (10.65) & 9.76 (10.17) & 10.03 (10.45) & 9.66 \\
 & \stt{$^3\Pi_u$} & Val. & 11.49 & 10.33 (10.93) & 12.00 & 11.44 (11.92) & 12.05 (12.71) & 11.55 (12.19) & 11.20 \\
 & \stt{$^3\Sigma_g^+$} & Ryd. & 12.40 & 11.86 (12.08) & 13.09 & 12.81 (13.02) & 13.00 (13.22) & 12.77 (12.99) & 12.30 \\
CO & \stt{$^3\Pi$} & Val. & 6.45 & 5.99 (6.32) & 6.81 & 6.68 (7.04) & 6.68 (7.02) & 6.57 (6.91) & 6.28 \\
 & \stt{$^3\Sigma^+$} & Val. & 8.54 & 8.09 (8.49) & 8.57 & 8.44 (8.75) & 8.60 (9.01) & 8.58 (8.99) & 8.44 \\
 & \stt{$^3\Delta$} & Val. & 9.34 & 8.89 (9.30) & 9.40 & 9.20 (9.52) & 9.41 (9.82) & 9.38 (9.80) & 9.26 \\
 & \stt{$^3\Sigma^-$} & Val. & 10.01 & 9.54 (9.89) & 10.25 & 9.97 (10.28) & 10.07 (10.42) & 10.04 (10.40) & 9.76 \\
 & \stt{$^3\Sigma^+$} & Ryd. & 10.83 & 10.52 (10.71) & 11.22 & 11.06 (11.25) & 11.18 (11.37) & 11.03 (11.22) & 10.47 \\
 & \stt{$^3\Sigma^+$} & Ryd. & 11.64 & 11.39 (11.47) & 12.09 & 11.88 (11.96) & 12.05 (12.13) & 11.90 (11.98) & 11.30 \\
C$_2$H$_2$ & \stt{$^3\Sigma_u^+$} & Val. & 5.75 & 5.65 (5.97) & 5.86 & 6.14 (6.49) & 5.77 (6.08) & 5.97 (6.29) & 5.52 \\
 & \stt{$^3\Delta_u$} & Val. & 6.58 & 6.49 (6.83) & 6.66 & 6.87 (7.23) & 6.60 (6.92) & 6.81 (7.15) & 6.40 \\
 & \stt{$^3\Sigma_u^-$} & Val. & 7.27 & 7.14 (7.47) & 7.38 & 7.52 (7.87) & 7.24 (7.56) & 7.46 (7.78) & 7.08 \\
C$_2$H$_4$ & \stt{$^3B_{1u}$} & Val. & 4.59 & 4.77 (5.09) & 4.97 & 5.51 (5.87) & 5.10 (5.41) & 5.29 (5.62) & 4.55 \\
 & \stt{$^3B_{3u}$} & Ryd. & 7.23 & 7.07 (7.16) & 7.39 & 7.42 (7.49) & 7.28 (7.37) & 7.45 (7.53) & 7.26 \\
 & \stt{$^3B_{1g}$} & Ryd. & 7.95 & 7.81 (7.89) & 8.17 & 8.18 (8.24) & 8.03 (8.09) & 8.19 (8.26) & 8.02 \\
CH$_2$O & \stt{$^3A_2$} & Val. & 3.46 & 3.49 (3.73) & 4.31 & 4.45 (4.69) & 4.62 (4.92) & 4.57 (4.87) & 3.57 \\
 & \stt{$^3A_1$} & Val. & 6.20 & 6.22 (6.70) & 6.36 & 6.83 (7.30) & 6.60 (7.07) & 6.71 (7.19) & 6.06 \\
 & \stt{$^3B_2$} & Ryd. & 6.39 & 6.18 (6.20) & 7.58 & 7.12 (7.12) & 7.31 (7.33) & 7.21 (7.23) & 7.08 \\
 & \stt{$^3B_2$} & Ryd. & 7.41 & 7.23 (7.33) & 8.54 & 8.22 (8.31) & 8.35 (8.44) & 8.26 (8.36) & 7.95 \\
 & \stt{$^3A_1$} & Ryd. & 7.40 & 7.23 (7.25) & 8.62 & 8.19 (8.21) & 8.35 (8.39) & 8.24 (8.27) & 8.11 \\
 & \stt{$^3B_1$} & Val. & 8.39 & 7.92 (8.54) & 9.09 & 8.96 (9.53) & 9.16 (9.82) & 9.04 (9.70) & 8.42 \\
 & \stt{$^3A_2$} & Ryd. & 8.01 & 7.31 (7.34) & 8.83 & 8.40 (8.41) & 8.53 (8.56) & 8.47 (8.49) & 8.69 \\
\midrule
MAE &  &  & 0.29 & 0.43 (0.43) & 0.43 & 0.38 (0.64) & 0.38 (0.66) & 0.36 (0.64) &  \\
MSE &  &  & 0.00 & $-$0.30 ($-$0.02) & +0.43 & +0.35 (+0.62) & +0.37 (+0.65) & +0.34 (+0.63) &  \\
RMSE &  &  & 0.36 & 0.54 (0.54) & 0.49 & 0.45 (0.73) & 0.47 (0.76) & 0.42 (0.72) &  \\
Max &  &  & 0.71 & 1.38 (1.35) & 0.80 & 0.96 (1.32) & 1.05 (1.51) & 1.00 (1.30) &  \\
\midrule
MAE &  & Val. & 0.21 & 0.30 (0.35) & 0.43 & 0.46 (0.85) & 0.39 (0.80) & 0.43 (0.85) &  \\
MSE &  & Val. & +0.19 & $-$0.09 (+0.31) & +0.43 & +0.46 (+0.85) & +0.39 (+0.80) & +0.43 (+0.85) &  \\
RMSE &  & Val. & 0.26 & 0.36 (0.44) & 0.47 & 0.53 (0.90) & 0.49 (0.87) & 0.48 (0.88) &  \\
Max &  & Val. & 0.57 & 0.87 (0.81) & 0.80 & 0.96 (1.32) & 1.05 (1.51) & 1.00 (1.30) &  \\
\midrule
MAE &  & Ryd. & 0.40 & 0.60 (0.53) & 0.44 & 0.26 (0.36) & 0.37 (0.47) & 0.27 (0.37) &  \\
MSE &  & Ryd. & $-$0.27 & $-$0.58 ($-$0.46) & +0.44 & +0.21 (+0.31) & +0.34 (+0.45) & +0.23 (+0.34) &  \\
RMSE &  & Ryd. & 0.46 & 0.71 (0.65) & 0.50 & 0.32 (0.43) & 0.44 (0.56) & 0.32 (0.43) &  \\
Max &  & Ryd. & 0.71 & 1.38 (1.35) & 0.79 & 0.59 (0.78) & 0.75 (0.92) & 0.60 (0.75) &  \\
\bottomrule
\end{tabular*}
\end{table*}

\begin{figure*}[hbt!]
    \centering
    \includegraphics[width=\textwidth]{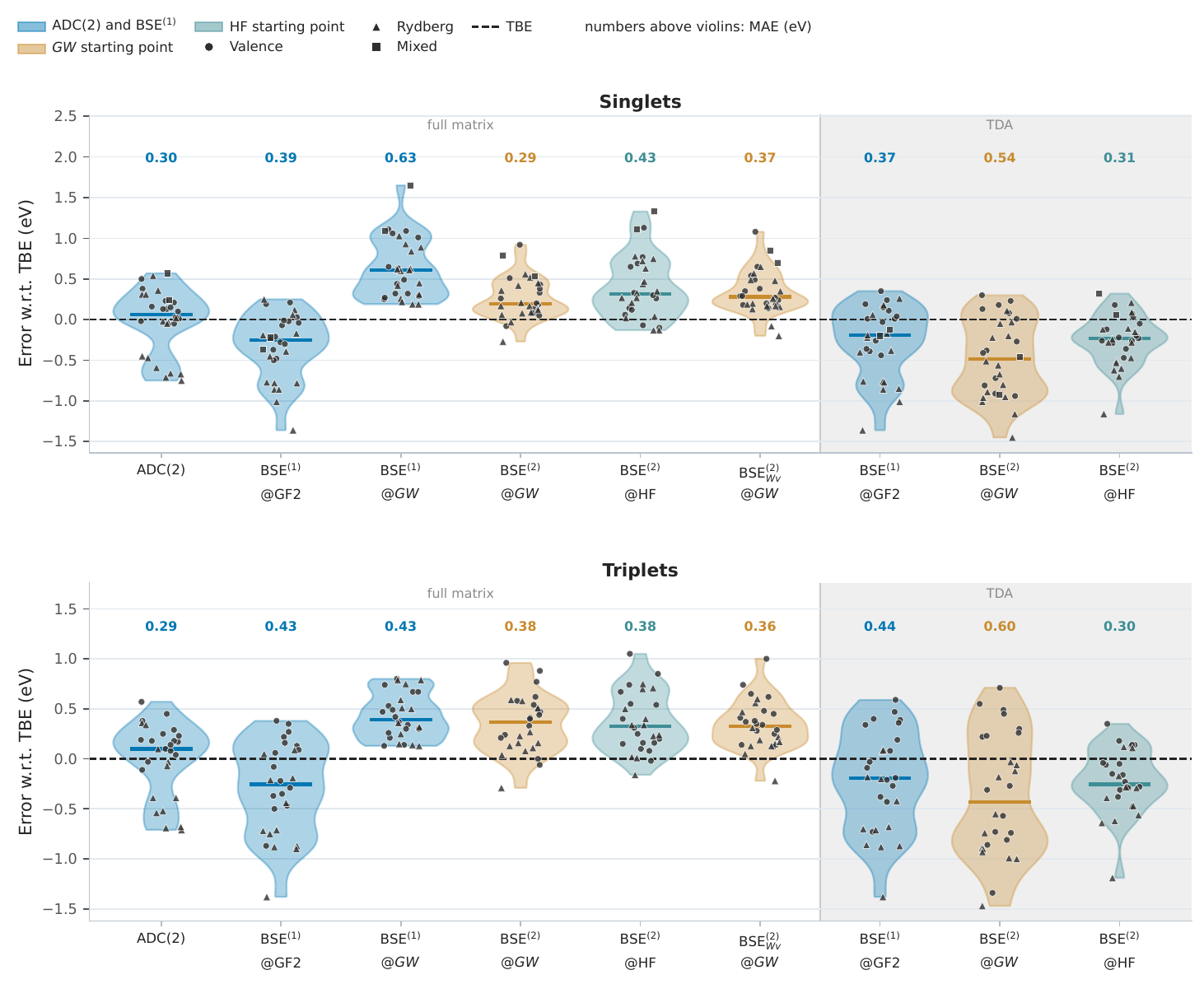}
    \caption{Error distributions for the excitation energies of Tables~\ref{tab:singlet_full}--\ref{tab:triplet_tda}, relative to the QUEST v1.2 TBEs. Positive values denote overestimation. Top: singlets. Bottom: triplets. Markers are individual states (circles: valence, triangles: Rydberg, squares: mixed), bars are medians, numbers are MAEs in eV, and the shaded band collects the TDA results.}
    \label{fig:violin}
\end{figure*}

\subsection{Overall accuracy}

For the 30 singlet states, BSE$^{(2)}$@$GW$ lowers the MAE from 0.63~eV for BSE$^{(1)}$@$GW$ to 0.29~eV, the smallest overall MAE in Table~\ref{tab:singlet_full} and on par with ADC(2) (0.30~eV). BSE$^{(2)}_{Wv}$@$GW$ (0.37~eV) and BSE$^{(2)}$@HF (0.43~eV) perform similarly. For the triplets, the three second-cycle variants are practically indistinguishable, with MAEs between 0.36 and 0.38~eV, compared to 0.43~eV for BSE$^{(1)}$@$GW$. For all three, the MSE is close to the MAE, so the remaining error is mainly a systematic overestimation of 0.3--0.4~eV. ADC(2) yields MAEs of 0.30 (singlets) and 0.29~eV (triplets) with a vanishing MSE, whereas BSE$^{(1)}$@GF2, with MAEs of 0.39 and 0.43~eV, is the only method that underestimates the excitation energies on average.

The MAE of BSE$^{(1)}$@$GW$ is in line with the 0.64 (singlets) and 0.41~eV (triplets) that Monino and Loos obtained for BSE@$G_0W_0$@HF on a smaller, overlapping set of excitations of the same seven molecules.~\cite{Monino2023} It exceeds the 0.1--0.3~eV usually quoted for $GW$+BSE.~\cite{Blase2020} Those figures were obtained mainly for the low-lying valence excitations of Thiel's set, which contains no Rydberg states, using ev$GW$ or $G_0W_0$ based on hybrid functionals with a large fraction of exact exchange.~\cite{Jacquemin2015,Bruneval2015}

In both sequences, the second cycle improves on the first. For the $GW$ sequence, it lowers 28 of the 30 singlet excitation energies, by 0.37~eV on average. The HF sequence shows an even larger improvement. TDHF is known to describe neutral excitations poorly and to develop triplet instabilities.~\cite{Cizek1967,Dreuw2005,Rangel2017} On the smaller set of Ref.~\citenum{Monino2023}, TDHF gives MAEs of 0.94~eV for singlets and 1.65~eV for triplets, about two and four times those of BSE$^{(2)}$@HF here. BSE$^{(2)}$@HF also yields a complete and stable set of triplet excitations.

\subsection{Role of screened interaction and QP energies}

The second-cycle variants differ only in the interaction entering $A^{(2)}$ and in the QP energies. We now analyze how each of these two parameters affects valence and Rydberg excitation energies.

We first isolate the effect of the screened interaction in the second-cycle kernel by comparing BSE$^{(2)}$@$GW$ with BSE$^{(2)}_{Wv}$@$GW$. Both methods use the same QP energies, but only the former builds the second-order kernel with the screened interaction $W_0$. For the singlets, screening lowers 26 of the 30 excitation energies, by at most 0.32~eV (CO $^1\Delta$), and reduces the MAE from 0.37 to 0.29~eV. For the triplets, the shifts range from $-0.18$ to $+0.24$~eV and average $+0.01$~eV, so the MAE is essentially unchanged (0.38~eV, compared with 0.36~eV).

Because BSE$^{(1)}$@GF2, BSE$^{(2)}$@HF, and BSE$^{(2)}_{Wv}$@$GW$ share the same kernel and differ only in their QP energies, comparing them isolates the effect of the vertex-corrected QP energies. Replacing the GF2 QP energies by the vertex-corrected ones substantially raises the excitation energies on average by 0.73 (singlets) and 0.67~eV (triplets) for BSE$^{(2)}$@HF, and by nearly the same amount for BSE$^{(2)}_{Wv}$@$GW$ (0.68 and 0.64~eV). The shift is larger for Rydberg states than for valence states. For BSE$^{(2)}$@HF, the Rydberg states shift by 0.83 (singlets) and 0.92~eV (triplets), almost twice as much as the valence states (0.48~eV for both). As shown in Figure~\ref{fig:violin}, this moves the error distribution as a whole, from underestimation to overestimation. The MSE therefore changes sign for both variants and both spin states, from about $-0.3$~eV to between $+0.34$ and $+0.40$~eV, while the MAE changes by at most 0.07~eV. The QP energies of the $GW$ sequence, however, also make the distribution markedly more compact. Compared with BSE$^{(2)}$@HF, BSE$^{(2)}_{Wv}$@$GW$ changes the MSE by only $-0.05$ (singlets) and $-0.02$~eV (triplets) but reduces the standard deviation of the signed errors, $(\mathrm{RMSE}^2-\mathrm{MSE}^2)^{1/2}$, from 0.38 to 0.26~eV for the singlets and from 0.29 to 0.24~eV for the triplets. 

Out of all methods considered here, including ADC(2), BSE$^{(2)}$@$GW$ and BSE$^{(2)}_{Wv}$@$GW$ give the lowest MAEs for Rydberg states (0.24--0.29~eV). This is expected, since the electron and hole of a Rydberg excitation mainly interact via the direct and long-ranged first-order interaction, for which $W_0 \approx v$ far from the molecule. This term is common to all methods considered here. Rydberg excitation energies are therefore dominated by the QP energies. These are more accurate in BSE$^{(2)}$@$GW$ and BSE$^{(2)}_{Wv}$@$GW$, which use the vertex-corrected QP energies of Ref.~\citenum{Forster2024} as input, than in BSE$^{(1)}$@GF2 and ADC(2), which only contain the GF2 self-energy. Consequently, ADC(2) and BSE$^{(1)}$@GF2 both underestimate Rydberg excitation energies, with relatively large MAEs of 0.40 and 0.55~eV for singlets and 0.40 and 0.60~eV for triplets. ADC(2), however, contains the dynamical GF2 line dressing that BSE$^{(1)}$@GF2 neglects, which is a plausible candidate to explain the differences between the two methods for Rydberg excitations.

In contrast, ADC(2) remains the best performer for valence excitations, closely followed by BSE$^{(1)}$@GF2. Overall, the vertex-corrected QP energies thus improve the Rydberg states but worsen the valence states. Our data support the hypothesis that the GF2-based methods benefit from error cancellation. Although the Rydberg states are too low, the valence states of BSE$^{(1)}$@GF2 are nearly unbiased, and with the dynamical correction (discussed in Sec.~\ref{sec:res-dynamical}) they are already too high (MSEs of $+0.20$ and $+0.31$~eV for singlets and triplets, and $+0.15$ and $+0.19$~eV for ADC(2)). This indicates that the second-order kernels of both methods underbind compact electron--hole pairs. Correcting the QP energies removes this cancellation.

\subsection{Dynamical corrections}
\label{sec:res-dynamical}

The dynamically corrected results are reported in parentheses in \Cref{tab:singlet_full,tab:triplet_full,tab:singlet_tda,tab:triplet_tda}. Since the dBSE and BSE energies share all static contributions, their difference, $\zeta_m\delta\Omega_m$, isolates the frequency-dependent part of $A^{(2)}$. This contribution is larger for valence than for Rydberg excitations. In the full eigenvalue problem, the correction blueshifts the valence singlets by 0.32--0.36~eV on average, depending on the method, compared with only 0.06--0.07~eV for the Rydberg singlets. For the triplets, the corresponding shifts are 0.40--0.42 and 0.10--0.11~eV, and the TDA gives similar values. On average, the dynamical correction thus acts almost exclusively as a valence-state blueshift. The renormalization factors follow the same split. They lie between 0.98 and 1.10 for all states and methods (Supporting Information), with averages of 1.06 for the valence and 1.01 for the Rydberg states. The renormalization therefore changes the corrections by only 0.01~eV on average and by at most 0.06~eV.

The correction also depends little on the QP energies and on the interaction in the second-cycle kernel. Between BSE$^{(1)}$@GF2, BSE$^{(2)}$@HF, and BSE$^{(2)}_{Wv}$@$GW$, the corrections to individual states differ on average by only 0.01--0.02~eV for both spin states. Replacing $v$ by $W_0$ in the second-cycle kernel matters more. The corrections of BSE$^{(2)}$@$GW$ and BSE$^{(2)}_{Wv}$@$GW$ differ on average by 0.05 (singlets) and 0.04~eV (triplets). Both effects are small, however, compared with the roughly 0.3~eV by which the correction shifts the valence states more than the Rydberg states in every method. Rebolini and Toulouse likewise found larger blueshifts for valence than for Rydberg states when adding the complete second-order kernel perturbatively within the TDA.~\cite{Rebolini2016} As discussed in the previous section, this observation is readily rationalized by the fact that Rydberg excitations are much less sensitive to the form of the second-order kernel than valence excitations.

Because the correction is essentially a fixed, character-dependent blueshift, it is not an unconditional improvement. At first sight, this is surprising, since dBSE energies are sometimes regarded as the more accurate estimates.~\cite{Ma2009,Loos2020b,Monino2023} However, dBSE@GF2 as implemented by Monino and Loos evaluates the same dynamical kernel but measures the correction from the conventional static limit (Eq.~\eqref{eq:second-cyclekernelbeforeeff} at $\omega=0$). As a result, their static BSE@GF2 energies overestimate the reference values and the correction lowers them. Our static BSE$^{(1)}$@GF2 energies, in contrast, underestimate the reference values (MSEs of $-0.33$ and $-0.30$~eV), and the correction raises them. This brings the MSEs close to zero ($-0.13$ and $-0.02$~eV) but hardly changes the MAEs, which decrease only from 0.39 to 0.36~eV (singlets) and remain at 0.43~eV (triplets). The underestimation is concentrated in the Rydberg states, which the correction shifts least, whereas the nearly unbiased valence states are pushed too high, raising the valence MAE from 0.21 to 0.22~eV (singlets) and from 0.30 to 0.35~eV (triplets).

For the BSE$^{(2)}$ variants, which already overestimate the excitation energies, the correction moves every valence state further from the TBE. The singlet MAEs increase from 0.29 to 0.45~eV for BSE$^{(2)}$@$GW$, from 0.43 to 0.61~eV for BSE$^{(2)}$@HF, and from 0.37 to 0.56~eV for BSE$^{(2)}_{Wv}$@$GW$, and the triplet MAEs from 0.36--0.38 to 0.64--0.66~eV. Within the TDA (Tables~\ref{tab:singlet_tda} and \ref{tab:triplet_tda}), where all static MSEs are negative, the correction lowers the singlet MAE of BSE$^{(2)}$@HF from 0.31 to 0.29~eV and leaves both MAEs of BSE$^{(2)}$@$GW$ nearly unchanged (0.54 to 0.55 and 0.60 to 0.61~eV), although it brings their MSEs closer to zero (from $-0.47$ to $-0.34$ and from $-0.37$ to $-0.12$~eV). It raises, however, both MAEs of BSE$^{(1)}$@GF2, as well as the triplet MAE of BSE$^{(2)}$@HF (from 0.30 to 0.35~eV). The blueshift thus can reduce a negative average bias, but it does not systematically improve the state-by-state accuracy.

Because ADC(2) contains a symmetrized, frequency-dependent version of Eq.~\eqref{eq:second-cyclekernel}, together with a static second-order term that accounts for the coupling block, comparing it with the full BSE$^{(1)}$@GF2 indicates what the static approximation neglects. On average, ADC(2) lies above the BSE$^{(1)}$@GF2 energies by about 0.3~eV for both the valence and the Rydberg states of either spin. For both spin states, the correction overshoots the valence gap ($+0.35$ and $+0.41$~eV) but recovers less than half of the Rydberg gap ($+0.06$ and $+0.11$~eV). The remaining differences, the dynamical corrections beyond the first order and the additional second-order terms contained in the ADC(2) secular equations, therefore appear to raise the Rydberg states by about 0.2~eV and to lower the valence states by about 0.05 (singlets) and 0.1~eV (triplets).

\begin{table}[hbt!]
\setlength{\aboverulesep}{0.25ex}\setlength{\belowrulesep}{0.4ex}
\providecommand{\stt}[1]{$\vphantom{X}$\smash{#1}}
\renewcommand{\stkone}[1]{#1}
\centering
\scriptsize
\setlength{\tabcolsep}{3pt}
\renewcommand{\arraystretch}{0.9}
\caption{Singlet excitation energies (in eV, aug-cc-pVTZ). All entries except the TBE are computed within the TDA, dynamically corrected (dBSE) values in parentheses. Statistics are errors with respect to the TBE. The two states of mixed valence--Rydberg character are included in the overall statistics but in neither subset.}
\label{tab:singlet_tda}
\begin{tabular*}{\columnwidth}{@{\extracolsep{\fill}}ll c c c c}
\toprule
\stkone{Mol.} & \stkone{Nat.} & \hBSEGFtwo & \hBSEGWmain & \hBSEHF & \hTBE \\
\midrule
HCl & Val. & 7.85 (7.97) & 7.46 (7.56) & 7.72 (7.85) & 7.84 \\
H$_2$O & Ryd. & 6.87 (6.97) & 6.62 (6.69) & 7.16 (7.27) & 7.63 \\
 & Ryd. & 8.49 (8.59) & 8.34 (8.42) & 8.80 (8.90) & 9.50 \\
 & Ryd. & 9.23 (9.34) & 9.04 (9.11) & 9.48 (9.60) & 9.99 \\
N$_2$ & Val. & 8.93 (9.31) & 8.41 (8.65) & 9.09 (9.49) & 9.32 \\
 & Val. & 10.07 (10.54) & 10.18 (10.75) & 9.62 (10.07) & 9.88 \\
 & Val. & 10.53 (10.98) & 10.37 (10.85) & 10.07 (10.51) & 10.29 \\
 & Ryd. & 12.57 (12.62) & 12.30 (12.28) & 12.82 (12.87) & 12.97 \\
 & Mix. & 12.89 (12.96) & 12.17 (12.04) & 13.15 (13.22) & 13.09 \\
 & Ryd. & 12.71 (12.81) & 12.58 (12.66) & 12.97 (13.08) & 13.09 \\
 & Mix. & 13.30 (13.74) & 12.96 (13.23) & 13.74 (14.17) & 13.42 \\
 & Ryd. & 14.27 (14.37) & 14.12 (14.20) & 13.90 (13.99) & 14.01 \\
CO & Val. & 8.59 (8.91) & 8.07 (8.31) & 8.55 (8.88) & 8.48 \\
 & Val. & 9.57 (9.94) & 9.21 (9.48) & 9.46 (9.83) & 9.93 \\
 & Val. & 9.81 (10.13) & 9.13 (9.26) & 9.71 (10.03) & 10.07 \\
 & Ryd. & 11.14 (11.16) & 10.74 (10.67) & 11.17 (11.18) & 10.96 \\
 & Ryd. & 11.58 (11.63) & 11.32 (11.37) & 11.61 (11.66) & 11.52 \\
 & Ryd. & 11.50 (11.57) & 11.16 (11.18) & 11.48 (11.54) & 11.72 \\
C$_2$H$_2$ & Val. & 7.13 (7.45) & 7.27 (7.64) & 6.89 (7.22) & 7.09 \\
 & Val. & 7.40 (7.72) & 7.44 (7.78) & 7.17 (7.49) & 7.43 \\
C$_2$H$_4$ & Ryd. & 7.20 (7.25) & 7.32 (7.36) & 7.09 (7.15) & 7.37 \\
 & Val. & 8.05 (8.28) & 8.13 (8.37) & 7.93 (8.16) & 7.90 \\
 & Ryd. & 7.85 (7.91) & 8.01 (8.07) & 7.76 (7.82) & 8.04 \\
CH$_2$O & Val. & 3.98 (4.22) & 3.70 (3.85) & 4.15 (4.40) & 3.97 \\
 & Ryd. & 6.36 (6.35) & 6.26 (6.21) & 6.60 (6.59) & 7.22 \\
 & Ryd. & 7.35 (7.41) & 7.32 (7.35) & 7.59 (7.64) & 8.12 \\
 & Ryd. & 7.37 (7.36) & 7.33 (7.30) & 7.62 (7.61) & 8.22 \\
 & Ryd. & 7.30 (7.32) & 7.21 (7.19) & 7.50 (7.51) & 8.66 \\
 & Val. & 8.75 (9.32) & 8.38 (8.75) & 8.90 (9.48) & 9.19 \\
 & Val. & 9.78 (10.03) & 9.56 (9.67) & 9.38 (9.60) & 9.43 \\
\midrule
MAE &  & 0.37 (0.40) & 0.54 (0.55) & 0.31 (0.29) &  \\
MSE &  & $-$0.26 ($-$0.07) & $-$0.47 ($-$0.34) & $-$0.24 ($-$0.05) &  \\
RMSE &  & 0.50 (0.51) & 0.67 (0.66) & 0.39 (0.38) &  \\
Max &  & 1.36 (1.34) & 1.45 (1.47) & 1.16 (1.15) &  \\
\midrule
MAE & Val. & 0.20 (0.31) & 0.41 (0.46) & 0.21 (0.19) &  \\
MSE & Val. & $-$0.03 (+0.31) & $-$0.27 (+0.01) & $-$0.17 (+0.17) &  \\
RMSE & Val. & 0.25 (0.38) & 0.52 (0.51) & 0.24 (0.23) &  \\
Max & Val. & 0.44 (0.69) & 0.94 (0.87) & 0.47 (0.43) &  \\
\midrule
MAE & Ryd. & 0.55 (0.51) & 0.64 (0.63) & 0.40 (0.36) &  \\
MSE & Ryd. & $-$0.48 ($-$0.42) & $-$0.62 ($-$0.60) & $-$0.36 ($-$0.31) &  \\
RMSE & Ryd. & 0.66 (0.63) & 0.77 (0.75) & 0.49 (0.46) &  \\
Max & Ryd. & 1.36 (1.34) & 1.45 (1.47) & 1.16 (1.15) &  \\
\bottomrule
\end{tabular*}
\end{table}
 
\begin{table}[hbt!]
\setlength{\aboverulesep}{0.25ex}\setlength{\belowrulesep}{0.4ex}
\providecommand{\stt}[1]{$\vphantom{X}$\smash{#1}}
\renewcommand{\stkone}[1]{#1}
\centering
\scriptsize
\setlength{\tabcolsep}{3pt}
\renewcommand{\arraystretch}{0.9}
\caption{Triplet excitation energies (in eV, aug-cc-pVTZ). All entries except the TBE are computed within the TDA, dynamically corrected (dBSE) values in parentheses. Statistics are errors with respect to the TBE.}
\label{tab:triplet_tda}
\begin{tabular*}{\columnwidth}{@{\extracolsep{\fill}}ll c c c c}
\toprule
\stkone{Mol.} & \stkone{Nat.} & \hBSEGFtwo & \hBSEGWmain & \hBSEHF & \hTBE \\
\midrule
H$_2$O & Ryd. & 6.57 (6.73) & 6.36 (6.49) & 6.86 (7.02) & 7.25 \\
 & Ryd. & 8.38 (8.51) & 8.25 (8.36) & 8.68 (8.83) & 9.24 \\
 & Ryd. & 8.83 (9.03) & 8.64 (8.80) & 9.07 (9.28) & 9.54 \\
N$_2$ & Val. & 8.29 (8.71) & 8.25 (8.72) & 7.84 (8.24) & 7.70 \\
 & Val. & 7.64 (8.09) & 7.28 (7.64) & 7.79 (8.25) & 8.02 \\
 & Val. & 9.27 (9.74) & 9.17 (9.69) & 8.82 (9.27) & 8.87 \\
 & Val. & 10.05 (10.49) & 10.15 (10.69) & 9.60 (10.02) & 9.66 \\
 & Val. & 10.47 (11.07) & 9.86 (10.20) & 10.92 (11.55) & 11.20 \\
 & Ryd. & 11.88 (12.10) & 11.75 (11.93) & 12.13 (12.34) & 12.30 \\
CO & Val. & 6.19 (6.52) & 5.97 (6.32) & 6.13 (6.46) & 6.28 \\
 & Val. & 8.25 (8.66) & 7.63 (7.87) & 8.15 (8.55) & 8.44 \\
 & Val. & 8.99 (9.40) & 8.40 (8.65) & 8.88 (9.29) & 9.26 \\
 & Val. & 9.55 (9.90) & 9.19 (9.44) & 9.45 (9.79) & 9.76 \\
 & Ryd. & 10.55 (10.75) & 10.35 (10.52) & 10.58 (10.77) & 10.47 \\
 & Ryd. & 11.39 (11.48) & 11.12 (11.21) & 11.42 (11.50) & 11.30 \\
C$_2$H$_2$ & Val. & 5.88 (6.21) & 5.97 (6.33) & 5.65 (5.97) & 5.52 \\
 & Val. & 6.59 (6.93) & 6.66 (7.03) & 6.36 (6.69) & 6.40 \\
 & Val. & 7.16 (7.49) & 7.30 (7.67) & 6.92 (7.25) & 7.08 \\
C$_2$H$_4$ & Val. & 5.02 (5.34) & 5.26 (5.64) & 4.90 (5.23) & 4.55 \\
 & Ryd. & 7.08 (7.17) & 7.20 (7.28) & 6.98 (7.06) & 7.26 \\
 & Ryd. & 7.82 (7.89) & 7.99 (8.06) & 7.73 (7.80) & 8.02 \\
CH$_2$O & Val. & 3.54 (3.79) & 3.30 (3.48) & 3.71 (3.97) & 3.57 \\
 & Val. & 6.40 (6.89) & 6.29 (6.73) & 6.24 (6.72) & 6.06 \\
 & Ryd. & 6.20 (6.22) & 6.08 (6.06) & 6.44 (6.46) & 7.08 \\
 & Ryd. & 7.25 (7.34) & 7.21 (7.29) & 7.48 (7.57) & 7.95 \\
 & Ryd. & 7.24 (7.26) & 7.18 (7.19) & 7.49 (7.51) & 8.11 \\
 & Val. & 7.99 (8.61) & 7.69 (8.17) & 8.14 (8.77) & 8.42 \\
 & Ryd. & 7.31 (7.34) & 7.22 (7.21) & 7.50 (7.53) & 8.69 \\
\midrule
MAE &  & 0.44 (0.49) & 0.60 (0.61) & 0.30 (0.35) &  \\
MSE &  & $-$0.22 (+0.06) & $-$0.37 ($-$0.12) & $-$0.22 (+0.06) &  \\
RMSE &  & 0.54 (0.59) & 0.70 (0.72) & 0.38 (0.42) &  \\
Max &  & 1.38 (1.35) & 1.47 (1.48) & 1.19 (1.16) &  \\
\midrule
MAE & Val. & 0.32 (0.46) & 0.55 (0.62) & 0.20 (0.33) &  \\
MSE & Val. & +0.03 (+0.44) & $-$0.15 (+0.22) & $-$0.08 (+0.33) &  \\
RMSE & Val. & 0.37 (0.56) & 0.63 (0.70) & 0.22 (0.38) &  \\
Max & Val. & 0.73 (1.01) & 1.34 (1.09) & 0.38 (0.68) &  \\
\midrule
MAE & Ryd. & 0.59 (0.53) & 0.66 (0.59) & 0.44 (0.38) &  \\
MSE & Ryd. & $-$0.56 ($-$0.45) & $-$0.66 ($-$0.57) & $-$0.40 ($-$0.30) &  \\
RMSE & Ryd. & 0.70 (0.64) & 0.79 (0.74) & 0.53 (0.48) &  \\
Max & Ryd. & 1.38 (1.35) & 1.47 (1.48) & 1.19 (1.16) &  \\
\bottomrule
\end{tabular*}
\end{table}

\subsection{The coupling block and instabilities}
\label{sec:res-tda}

Tables~\ref{tab:singlet_tda} and~\ref{tab:triplet_tda} repeat the comparison within the TDA. For the BSE$^{(2)}$ variants, the TDA is applied in all cycles, so it changes not only the second-cycle BSE but also the vertex-corrected QP energies that enter it. For BSE$^{(1)}$@GF2, where it only removes the coupling block, the TDA changes the MAE by at most 0.02~eV (from 0.39 to 0.37~eV for singlets and from 0.43 to 0.44~eV for triplets) but raises individual valence states by up to 0.65~eV. For the BSE$^{(2)}$ variants, the TDA has a much larger effect. It lowers 29 (BSE$^{(2)}$@HF) and 25 (BSE$^{(2)}$@$GW$) of the 30 singlet and 27 of the 28 triplet excitation energies, by 0.59--0.74~eV on average, so that every MSE changes sign, from $+0.40$ to $-0.24$~eV (singlets) and from $+0.37$ to $-0.22$~eV (triplets) for BSE$^{(2)}$@HF, and from $+0.26$ to $-0.47$~eV and from $+0.35$ to $-0.37$~eV for BSE$^{(2)}$@$GW$. For BSE$^{(2)}$@HF, this improves the MAE from 0.43 to 0.31~eV (singlets) and from 0.38 to 0.30~eV (triplets), mainly through the valence states (from 0.37 to 0.21 and from 0.39 to 0.20~eV). For BSE$^{(2)}$@$GW$, the lowering overshoots. The MAE increases from 0.29 to 0.54~eV and from 0.38 to 0.60~eV, mainly because the Rydberg MAE rises from 0.24 to 0.64 and from 0.26 to 0.66~eV. Since BSE$^{(2)}$@HF shares its kernel with BSE$^{(1)}$@GF2, the large TDA effect on the BSE$^{(2)}$ variants probably stems mainly from the QP energies of the TDA in the first cycle.

To assess the role of the pp channel, we repeated the calculations with its contribution switched off. This amounts to omitting the second term of Eq.~\eqref{eq:effectivekernelsecondorder}, or, equivalently, the last two terms in Eq.~\eqref{eq:second-cyclekernel}. Several roots then acquire a substantial de-excitation character, with $\sum_{ia}Y_{ia}^{2} > 0.3$. These states approach an instability and therefore yield less reliable excitation energies. We consequently do not include these results in the main text. The corresponding data are reported in the Supporting Information. The pp terms describe the repulsion between two excited electrons and between two holes. In the HF sequence, where the pp terms only contain the bare interaction, their removal lowers the excitation energies and produces several triplet instabilities. In the $GW$ sequence, they contain a weaker interaction. Their removal therefore changes most excitation energies only slightly, and within the TDA it even raises them. It nevertheless pushes several singlets toward an instability. This mechanism may also explain why BSE$^{(2)}$@HF yields a complete and stable set of triplet excitations, even though its first-cycle TDHF kernel develops triplet instabilities. TDHF contains only ph contributions and therefore lacks this repulsive pp contribution. %
\section{Conclusions}
\label{sec:conclusions}

We have truncated the iterative sequence of Eqs.~\eqref{eq:nthorderHedin1}--\eqref{eq:nthorderHedin4} after the construction of $L^{(2)}$, starting from either the HF or the $G_0W_0$@HF self-energy in a fixed RHF orbital basis, and benchmarked the resulting excitation energies against the TBEs of QUEST v1.2, for singlets and triplets, within and beyond the TDA, and with and without a perturbative treatment of the dynamical kernel. In the full problem, with the effective second-order kernel evaluated at zero frequency, the second cycle lowers the MAE in both sequences. For the $GW$ sequence, whose first cycle is standard $GW$+BSE with a screened kernel, the MAE decreases from 0.63 to 0.29~eV for singlets and from 0.43 to 0.38~eV for triplets. For the HF sequence, whose first cycle, TDHF, describes neutral excitations poorly and is prone to triplet instabilities,~\cite{Cizek1967,Dreuw2005,Rangel2017} BSE$^{(2)}$@HF yields MAEs of 0.43 and 0.38~eV and a complete and stable set of triplet excitations.

How screening enters the second cycle requires care. Assigning $W_0$ to the exchange part of the $GW$ self-energy retains the screened exchange term in $A^{(1)}$, whose second-order part is already contained in $A^{(2)}$ and is therefore counted twice (App.~\ref{app:doublecounting}). The resulting double counting strongly blueshifts the valence excitations. With the bare exchange instead, $W_0$ enters the second-cycle kernel only through its exchange-type terms. Dressing these terms with $W_0$ improves the singlets moderately but leaves the triplets essentially unchanged. Whether screening the exchange term in $A^{(1)}$ to all orders without double counting would improve on BSE$^{(2)}$@$GW$ remains open.

The QP energies shift the neutral excitation energies considerably more. In the full eigenvalue problem, replacing the GF2 QP energies by vertex-corrected ones raises the excitation energies by about 0.7~eV on average, considerably more for Rydberg than for valence states. This improves the Rydberg states but degrades the valence states, and changes the MAE by at most 0.07~eV. Vertex-corrected QP energies are thus not decisively better than GF2 ones as input to the second-order kernel. Rebolini and Toulouse~\cite{Rebolini2016} likewise concluded, for a set of small molecules, that the accuracy of the second-order kernel depends crucially on the QP correction of the HF orbital energies. Neglecting the coupling block $B$ (TDA) hardly changes the MAE of BSE$^{(1)}$@GF2. For the BSE$^{(2)}$ variants, where the TDA is applied in all cycles, it lowers nearly all excitation energies by 0.59--0.74~eV on average. For BSE$^{(2)}$@HF, this reduces the MAEs to 0.31 and 0.30~eV, on par with ADC(2)~\cite{Loos2020d} (0.30 and 0.29~eV) and CC2~\cite{Christiansen1995,Loos2018b} (0.30~eV for both) for the same states. For BSE$^{(2)}$@$GW$, however, the lowering overshoots.

The second cycle, solved with a Davidson algorithm, has the same $\mathcal{O}(N^5)$ scaling as ADC(2) and CC2, but the vertex-corrected QP energies require an $\mathcal{O}(N^6)$ first cycle. The second-cycle variants of the $GW$ sequence outperform ADC(2) and CC2 only for the Rydberg states (0.24--0.29 versus 0.34--0.40~eV) due to the more accurate QP energies they use. ADC(2) remains preferable for valence excitations.

The present results should be interpreted in light of several approximations made in the construction of the scheme. In addition to the truncation, we neglect $\delta W/\delta G$ in the first-cycle kernel of the $GW$ sequence, replace $L^{(1)}$ by $L^{(2)}_0$ in the second-cycle functional derivative, take the static limit of the kernel, and update only the eigenvalues of $G$. The last may not be innocuous, since virtual HF orbitals in particular are much too diffuse, especially in small molecules, compared to the true Dyson orbitals.\cite{Baerends2013, Amati2020} The Green's function formalism comes with the major advantage that it provides a systematic framework to improve over HF orbitals.

The consistent improvement of the second cycle over the first suggests that higher cycles are worth pursuing. They follow from iterating Eqs.~\eqref{eq:nthorderHedin1}--\eqref{eq:nthorderHedin4} further, which will likely require automated code generation. Updating the functional form of the kernel in each cycle, rather than only its numerical values, is what Mejuto-Zaera and Vl\v{c}ek termed functional self-consistency. They analyzed this for the self-energy of the Hubbard dimer and showed that it can extend the validity of perturbative approximations based on Hedin’s equations toward the high-interaction limit.~\cite{Mejuto-Zaera2022a}

The dynamical second-order kernel used here contains all second-order diagrams that are irreducible in the ph channel, so that, unlike in $GW$+BSE, the vertex generated by the BSE is crossing symmetric through second order in $v$. This symmetry is broken by the static limit and by the absence of the transverse ph and pp diagrams beyond second order. Since the present static scheme is thus neither conserving nor crossing symmetric, it remains open which of the two properties matters more for the accuracy of excitation energies. One immediate extension is to resum the transverse particle--hole and particle--particle channels beyond second order, as in parquet approaches.~\cite{Krien2021,Marie2025}

A second immediate extension concerns the frequency dependence of the kernel. The perturbative correction considered here is nearly independent of the QP energies and acts almost exclusively as a blueshift of the valence states, by 0.3--0.4~eV on average. This reduces the negative bias of BSE$^{(1)}$@GF2 and of the TDA results but increases the errors of the full-matrix BSE$^{(2)}$ variants. Upfolding the dynamical kernel~\cite{Bintrim2022,Monino2023} or averaging it over frequencies with suitable weight functions, as proposed for the projection between channels in the frequency-dependent functional renormalization group,~\cite{Yirga2021} would provide a more robust treatment. Such a frequency-consistent treatment will also be needed to include the transverse channel, since static evaluations of the vertex tend to overestimate the coupling between channels.~\cite{Husemann2012,Vilardi2017} 
\appendix
\section{Double counting}
\label{app:doublecounting}

In BSE$^{(2)}$@$GW$, the screened interaction $W_0$ enters the second-cycle eigenvalue problem in two ways. First, it enters the QP energies of $G^{(1)}$, which are obtained from $\Sigma^\text{BSE}@L^\text{BSE}$, and hence the diagonal of $A^{(1)}$ and the denominators of Eq.~\eqref{eq:second-cyclekernelbeforeeff}. Second, it enters $A^{(2)}$ and $B^{(2)}$ through the exchange-type terms, because $\Xi^{(1)}$ appears in $\Sigma^{(1)}_c$ in Eq.~\eqref{eq:funcderiv}. With the partition of Eq.~\eqref{eq:sigmaGWex}, $W_0$ would enter in a third way. The first-order blocks would contain the screened terms $-(ij|W_0|ab)$ and $-(ib|W_0|ja)$ instead of the bare ones in Eq.~\eqref{eq:matrix-elements}. It is this third route that causes the double counting.
 
To make this explicit, we expand the RPA screened interaction in powers of the bare interaction,
\begin{equation}
    W = v + vP_0v + O(v^3) \,,
\end{equation}
where, for real orbitals and a closed-shell reference, 
\begin{align} \label{polarizability_real}
P_{0}(\br_1,\br_2,\omega)= & \, 2\sum_{kc}\varphi_c(\br_1)\varphi_c(\br_2)\varphi_k(\br_2)\varphi_k(\br_1) \nn\\ 
&\quad \times\left\{\frac{1}{\omega - (\varepsilon_c - \varepsilon_k) + i\eta} - \right. \nn\\ 
&\qquad \qquad\left. \frac{1}{\omega + (\varepsilon_c - \varepsilon_k) - i\eta} \right\} \,.
\end{align}
Evaluating the screened interaction at $\omega=0$ and inserting Eq.~\eqref{polarizability_real} gives
\begin{align}
  &-(ij|W_0|ab) \nn \\
  &\quad= -(ij|ab) + 4\int d(\br_1 \br_2 \br_3 \br_4)\varphi_i(\br_1)\varphi_j(\br_1)v(\br_1,\br_3) \nn \\
  &\qquad \times\sum_{kc}\frac{\varphi_c(\br_3)\varphi_c(\br_4)\varphi_k(\br_4)\varphi_k(\br_3)}{\varepsilon_c-\varepsilon_k} \nn\\
  &\qquad \times v(\br_4,\br_2)\varphi_a(\br_2)\varphi_b(\br_2) + O(v^3) \nn \\
  &\quad=-(ij|ab) + 4\sum_{kc}\frac{(ab|kc)(kc|ij)}{\varepsilon_c-\varepsilon_k} + O(v^3)\,,
  \label{eq:app-Wexp}
\end{align}
where the factor 4 combines the spin factor 2 of Eq.~\eqref{polarizability_real} with the two pole terms, which are equal at $\omega=0$. At $\omega=0$, the fourth term of each $kc$ sum in Eq.~\eqref{eq:second-cyclekernelbeforeeff} gives $2(ab|kc)(kc|ij)/(\varepsilon_c-\varepsilon_k)$, so the two sums together reproduce this second-order term.

Thus, with the partition of Eq.~\eqref{eq:sigmaGWex}, $A^{(1)}$ and $A^{(2)}$ contain the same second-order integrals with the same sign. They differ only in the energy denominator appearing in the effective kernel. Since
\begin{equation}
    \frac{\varepsilon_c-\varepsilon_k}{\varepsilon_b-\varepsilon_i+\varepsilon_c-\varepsilon_k}\,,\;
    \frac{\varepsilon_c-\varepsilon_k}{\varepsilon_a-\varepsilon_j+\varepsilon_c-\varepsilon_k} \in (0,1) \,,
\end{equation}
the contribution from the screened interaction in $A^{(1)}$ is, term by term, larger in magnitude than the corresponding fourth terms of both $kc$ sums in Eq.~\eqref{eq:second-cyclekernel}. Including the screened first-order terms raises the valence excitation energies by more than 1~eV (Supporting Information). This shift contains the doubly counted second-order contribution as well as the higher-order screening of these terms. 
\section*{Data and Software Availability Statement}

The data underlying this study are available within the published article and its Supporting Information.

\begin{suppinfo}
The Supporting Information contains the complete singlet and triplet excitation energies in the aug-cc-pVTZ basis set, including the double-counting variant, and calculations in which the pp channel is omitted, together with the renormalization factors $\zeta_m$ of the dynamical correction. It also provides the molecular geometries used for the AMS calculations.
\end{suppinfo}

\section*{Author Declarations}
The authors declare no conflict of interest.

\section*{Acknowledgments}
We thank Anna Kauch for insightful discussions on the frequency dependence of the kernel and Lucas Visscher for pointing out the dependence on the chemical potential of the kernel. The authors acknowledge funding by the Netherlands Organization for Scientific Research (NWO) under grant agreement OCENW.M.24.025. AF acknowledges funding through a VENI grant from NWO under grant agreement VI.Veni.232.013.

\bibliography{all}

\begin{tocentry}
\includegraphics[width=\textwidth]{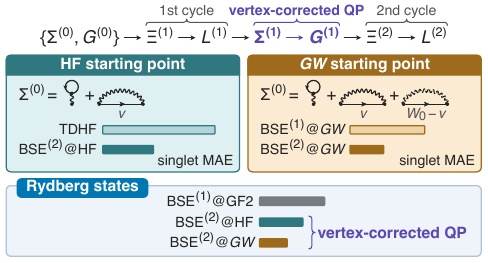}
\end{tocentry}

\end{document}